\documentclass[showpacs,aps,prd,preprintnumbers,amsmath,amssymb,superscriptaddress,groupedaddress,nofootinbib]{revtex4}
\usepackage{graphicx,subfigure,multirow,xcolor}
\usepackage{dcolumn,lipsum}% Align table columns on decimal point
\usepackage{bm}% bold math
\usepackage{natbib}
\usepackage[mathscr]{eucal}
\usepackage{doi}%<----------
\usepackage{soul}%this package is for text strike out
\usepackage{slashed} 
\usepackage[toc,page]{appendix}
\usepackage{float}
\usepackage{amsmath,amssymb}
\usepackage{subfigure}
\usepackage{float}
\usepackage{slashed}
\usepackage{slashbox}
\usepackage{multirow}
\usepackage{shorthand}
\makeatletter

\def\vev#1{\left\langle #1\right\rangle}

\begin{document}
\title{Probing the inert doublet model using jet substructure
	with a multivariate analysis }
%\date{\today}
%%%%%%%%%   Authors   %%%%%%%%%%%%
\author{Akanksha Bhardwaj}
\email{akanksha@prl.res.in}
\affiliation{Physical Research Laboratory (PRL), Ahmedabad - 380009, Gujarat, India}
\affiliation{Indian Institute of Technology, Gandhinagar-382424, Gujarat, India}
\author{Partha Konar}
\email{konar@prl.res.in}
\affiliation{Physical Research Laboratory (PRL), Ahmedabad - 380009, Gujarat, India}
\author{Tanumoy Mandal}
\email{tanumoy.mandal@physics.uu.se}
\affiliation{Department of Physics and Astronomy, Uppsala University, Box 516, SE-751 20 Uppsala, Sweden}
\author{Soumya Sadhukhan}
\email{physicsoumya@gmail.com}
\affiliation{Department of Physics and Astrophysics, University of Delhi, Delhi 110 007, India}

\begin{abstract}
	We explore the challenging but phenomenologically interesting hierarchical mass spectrum of the Inert Doublet Model where relatively light dark matter along with much heavier scalar states can fully satisfy the constraints on the relic abundance and also fulfill other theoretical as well as collider and astrophysical bounds. To probe this region of parameter space at the LHC, we propose a signal process that combines up to two large radius boosted jets along with substantial missing transverse momentum. Aided by our intuitive signal selection, we capture a hybrid process where the di-fatjet signal is significantly enhanced by the mono-fatjet contribution with minimal effects on the SM di-fatjet background. Substantiated by the sizable mass difference between the scalars, these boosted jets, originally produced from the hadronic decay of massive vector bosons, still carry the inherent footprint of their root. These features implanted inside the jet substructure can provide additional handles to deal with a large background involving QCD jets. We adopt a multivariate analysis using boosted decision tree to provide a robust mechanism to explore the hierarchical scenario, which would bring almost the entire available parameter space well within reach of the 14 TeV LHC runs with high luminosity.
\end{abstract}

\keywords{12.60.Jv,13.85.-t,95.35.+d}

%\allowdisplaybreaks

\maketitle
\flushbottom

%%%%%%%%%%%%%%%%%%%%%%%%%%%%%%%%%%%%%%%%%%%%%%%%%%%%%%
%%%%%%%%%%%%%%%%%%%%%%%%%%%%%%%%%%%%%%%%%%%%%%%%%%%%%%
\section{Introduction}
\label{sec:intro}
%%%%%%%%%%%%%%%%%%%%%%%%%%%%%%%%%%%%%%%%%%%%%%%%%%%%%%

The Standard Model (SM) of particle physics encapsulates our knowledge of fundamental interactions of the particle world with all its glory. Until now, apart from a few minor exceptions, the SM is in perfect agreement with all the high energy collider
experiments like the Large Hadron Collider (LHC) experiments at CERN. 
The reputation of the SM, being a complete theory, gets tarnished when it cannot explain the presence of tiny yet nonzero masses of the neutrinos that are already established in neutrino oscillation experiments. 
Observations of cosmic microwave background radiation in various experiments unambiguously establish that 26\% of the energy budget of our Universe is made up of an inert, stable component, termed as the `dark matter' (DM). The SM does not contain any particle that can satisfy the observed density of the DM, along with explaining its other properties. The inert doublet model (IDM) is proposed~\cite{Barbieri:2006dq, Cirelli:2005uq} as a minimal extension of the SM that can provide an inert weakly interacting DM candidate, stabilized by the discrete symmetry of the model. The SM is extended with an extra 
$\mathrm{SU}(2)_L$ scalar doublet which is odd under a discrete $\mathbb{Z}_2$ symmetry, and thus stabilizes the lightest neutral scalar of the model to be an ideal DM candidate. The neutrino mass can also be arranged in this set-up 
by introducing lepton portals~\cite{Borah:2017dqx}.

Inspection of the dark sector of the IDM, as done in Refs.~\cite{Ilnicka:2015jba, Belyaev:2016lok}, reveals that only small islands of parameter space can satisfy the full relic density of the DM dictated by the WMAP and the Planck results. Only light DM with a mass close to half of the Higgs mass can produce full relic density through the resonant Higgs portal annihilation. Even so, this requires a large mass difference between the DM with the additional beyond the SM (BSM) scalars in the model. Otherwise, the coannihilation effects in a degenerate mass spectrum reduce the relic density to under-produce DM. Another part of the parameter space where one can explain the full DM relic density is for the DM mass, $m_{\rm DM} \gtrsim 550$ GeV and even that is possible for an extremely degenerate BSM scalar mass spectrum. Among the various DM scenarios discussed above, heavy DM with a hierarchical spectrum can accommodate only about a few percent of the observed relic density and thus makes this scenario uninteresting to probe at the LHC.

One of the earlier collider studies of the IDM is performed in Ref.~\cite{Cao:2007rm}.
The dilepton and trilepton signatures at the LHC originating from the IDM have been investigated in 
Refs.~\cite{Dolle:2009ft,Miao:2010rg}. Although the degenerate heavy DM scenario can provide full relic density, the inertness of the model leads to kinematic suppression in heavy DM production at colliders and therefore, makes the signal very weak. Moreover, 
detection of the soft decay products from such a compressed mass spectrum remains challenging due to poor signal efficiency. This scenario is probed using charged track signatures by the CMS Collaboration~\cite{CMS:2014gxa}. 

Among the two light DM scenarios in this model, the degenerate BSM scalar mass spectrum can satisfy only about 10\% of the observed DM relic density. Nonetheless, this case is probed at the LHC through the mono jet search in Ref.~\cite{Belyaev:2018ext}. 
We motivate our framework with the light DM along with the hierarchical mass spectrum where the full DM relic density is achieved. 
Although very challenging due to tiny production cross sections of the unstable heavy BSM scalars at the LHC, 
their significant mass differences with the DM candidate give rise to interesting signal topology 
characterized by two boosted jets along with large missing transverse energy (MET) from the DM production. 
This gives us a scope to employ sophisticated multivariate analysis equipped with
jet substructure variables to isolate the signal from an overwhelmingly large SM background. 
The search for BSM scalars for this case is studied in the dijet plus MET channel, in a recent 
study~\cite{Poulose:2016lvz}. All these searches do not exhibit bright discovery potential even with the highest possible 
LHC luminosity. We look to explore a suitable discovery potential of this scenario, in this paper.

To reiterate the scenario under consideration, the heavier BSM scalars (a pseudoscalar $A$ and charged Higgs $H^\pm$) 
reside in the mass range ($\sim$ 250 - 700 GeV) much higher than the small mass window (55 - 80 GeV) where the DM candidate can lie. 
Hence, the hierarchical mass difference is large enough for these heavy scalars to decay dominantly 
to vector bosons ($V=\{W,Z\}$), which in turn are sufficiently boosted and corresponding hadronic deposits 
at the calorimeter behave like large radius fatjets, characterized by the jet radius, $R \sim p_T/2 m_V \gtrsim 0.8$.  
Accompanied with large MET acquired by the undetected pair of DM particles, the presence of these fatjets in the signal brings
additional variables like fatjet mass ($M_{J}$) and subjettiness ($\tau_{21}$) which carry the characteristics of 
boosted $W/Z$ decay. These observables are perceived as the ones that can distinguish well between the signal and 
the background, dominantly coming from the SM $V+$jets, as only a tiny fraction of QCD-jets mimic as boosted jets. Still, when the overwhelmingly large cross section of the background is pitted against suppressed signal cross section due to inertness of the model, even a tiny fraction can overshadow the signal to deny a significant discovery potential.

There is a possibility of mono-fatjet signal topology \cite{Khachatryan:2016mdm,Aaboud:2018xdl} with roughly one order of magnitude higher cross section than the
di-fatjet one. In this case, although we have a more significant cross section, the corresponding background becomes uncontrollably
large. Therefore, the mono-fatjet topology alone is not sufficient to achieve discovery significance. The di-fatjet
topology, on the other hand, also by itself is not enough to produce discovery significance due to tiny production cross
section.
In this paper, we propose a hybrid topology where the signal selections are designed aiming the di-fatjet topology but 
also allow a substantial fraction of the mono-fatjet signal. In doing this, we not only gain in the signal but at the
same time, the enormous mono-fatjet background can also be tamed down.

A probe of the IDM using a cut-based analysis (CBA) in the di-fatjet plus MET channel, has failed to reach discovery significance of $5\sg$ in any of our chosen benchmark points. From the nature of event distribution profiles, 
it is observed that the two variables viz. $M_{J}$ and $\tau_{21}$ are very powerful to separate tiny signal from the enormous SM background. The discovery significance in CBA is still elusive even if these jet substructure variables are used to the hilt. A multivariate analysis (MVA) in general performs better than a CBA, if appropriate kinematic variables are used in the analysis. 
So, a sophisticated MVA involving jet sub-structure variables quoted 
above is imperative to achieve better discovery potential in the IDM. 
With the events selected only after baseline cuts (defined later), the signal is still too tiny compared to the background to train the MVA set-up. 
Therefore, the baseline selection criteria should be accompanied by stronger selection cuts at the baseline level to cut down the large background without harming the signal too much before passing events to MVA. 
These cuts should be chosen optimally; otherwise, if they are very similar to the one used in CBA, eventually reduce the predictive power of MVA. 
Finally, our selection cuts are designed in a way such that it allows signal to consist of two high-$p_T$ fatjets.  Along with the pure di-fatjet signal, substantial contribution comes from the events with mono-fatjet that mimics the signal. 
This admixture significantly increases the signal cross section but simultaneously bring in some extra background processes in the picture. 
We perform an MVA analysis coupled with jet substructure variables to achieve improved signal vs background discrimination, which could not be realised in the CBA.
It helps us to reach a significantly higher LHC discovery potential in the di-fatjet plus MET channel of the IDM.  

This paper is organised as follows. In Sec.~\ref{sec:idm} we briefly discuss the IDM, outlining its scalar sector.  
Next, in Sec.~\ref{cons-idm}, we invoke all the possible theoretical, collider and astrophysical constraints applicable to the IDM, to ascertain the viability of hierarchical BSM scalar sector along with the presence of a light DM.
Subsequently, in Sec.~\ref{sec:BP}, we discuss four possible DM scenarios depending on the DM mass and its mass differences to the
other BSM scalars to motivate our choice of benchmark points. To define our analysis set-up, we list the possible IDM processes contributing to our signal process, di-fatjet plus MET channel in Sec.~\ref{coll}. 
We also discuss all possible SM backgrounds for this channel. At this point, we present our sample benchmark points covering our region of interest, which is also consistent with all the discussed constraints. In Sec.~\ref{sec:CBA}, we first use the baseline cuts and then introduce two fatjet specific observables and study how these perform to increase the signal vs background ratio and obtain the LHC reach for all the benchmark points. In Sec.~\ref{sec:mva}, we improve our probe using MVA to recast the signal vs background numbers with non-rectangular cuts and therefore, having better sensitivity for the LHC search. Finally, we summarize our results and conclude in Sec.~\ref{results}.

%%%%%%%%%%%%%%%%%%%%%%%%%%%%%%%%%%%%%%%%%%%%%%%%%%%%%%
\section{Inert Doublet Model}
\label{sec:idm}
%%%%%%%%%%%%%%%%%%%%%%%
We first discuss the traditional IDM where one adds an additional $\mathrm{SU}(2)_L$ complex scalar doublet $\Phi_2$ apart from the SM Higgs doublet $\Phi_1$, which are, respectively, odd and even under a discrete $\mathbb{Z}_2$ symmetry, {\emph i.e.} $\Phi_1 \rightarrow \Phi_1$, $\Phi_2 \rightarrow -\Phi_2$. The most general scalar potential that respects the electroweak symmetry $\mathrm{SU}(2)_L \otimes \mathrm{U}(1)_Y \otimes \mathbb{Z}_2$ of the IDM can be written as~\cite{Belyaev:2016lok},
\begin{align}
V(\Phi_1,\Phi_2)&=\mu_1^2 \Phi_1^\dagger \Phi_1 + \mu_{2}^2 \Phi_2^\dagger\Phi_2 +\lambda_1(\Phi_1^\dagger \Phi_1)^2    
+\lambda_2(\Phi_2^\dagger \Phi_2)^2
+\lambda_3 \Phi_1^\dagger\Phi_1 \Phi_2^\dagger\Phi_2 \nn\\ 
&+\lambda_4 \Phi_1^\dagger\Phi_2 \Phi_2^\dagger\Phi_1+\dfrac{\lambda_5}{2} \left[(\Phi_1^\dagger\Phi_2)^2 + H.c. \right],  
\label{v2hdm}
\end{align}
where $\Phi_1$ and $\Phi_2$ both are hypercharged, $Y=+1$, and can be written as
\begin{equation}
\label{Higgs}
\Phi_1 	=
	\begin{pmatrix}
		G^+\\ \displaystyle\frac{v +h +i G^0}{\sqrt{2}} 
	\end{pmatrix}, \ \  
\Phi_2 = \begin{pmatrix} H^+\\  \displaystyle\frac{H+iA}{\sqrt 2} \end{pmatrix}.
\end{equation}
 Here $h$ is the SM Higgs with $G^{+}, G^{0}$ being the charged and neutral Goldstone bosons, respectively. The charged scalar $H^+$ is present in $\Phi_2$, along with the neutral scalars, $H, A$, respectively, being CP-even and CP-odd. For the vacuum expectation values (VEVs) of the two doublets, we adopt the notation $\vev{\Phi_1}=v / \sqrt{2}$, $\vev{\Phi_2}= 0 $, keeping in mind the exact nature of the $\mathbb{Z}_2$ symmetry. 
The zero VEV of $\Phi_2$ is responsible for the inertness of this model. Since all the SM fermions are even under $\mathbb{Z}_2$, the new scalar doublet does not couple to the SM fermions and thus having no fermionic interactions. The scalar-gauge boson interactions originate through the kinetic term of the two doublets 
\begin{equation}
\label{HiggsKin}
{\cal L}_{kin} = (D_{\mu}\Phi_1)^{\dagger}  (D^{\mu} \Phi_1) + (D_{\mu}\Phi_2)^{\dagger}  (D^{\mu} \Phi_2).
\end{equation}
All parameters in the scalar potential are assumed to be real in order to keep the IDM CP-invariant. 

Here, the electroweak symmetry breaking takes place through the SM doublet $\Phi_1$ getting a VEV, and after this, the masses of the physical scalars at tree level can be written as
\begin{align}
m_h^2 &= 2\lambda_1 v^2 ,\nonumber\\
m_{H^\pm}^2 &= \mu_2^2 + \frac{1}{2}\lambda_3 v^2 , \nonumber\\
m_{H}^2 &= \mu_2^2 + \frac{1}{2}(\lambda_3+\lambda_4+\lambda_5)v^2=m^2_{H^\pm}+
\frac{1}{2}\left(\lambda_4+\lambda_5\right)v^2  , \nonumber\\
m_{A}^2 &= \mu_2^2 + \frac{1}{2}(\lambda_3+\lambda_4-\lambda_5)v^2=m^2_{H^\pm}+
\frac{1}{2}\left(\lambda_4 - \lambda_5  \right)v^2.
\label{mass_relation}
\end{align}
 Here, $m_h$ is the SM-like Higgs boson mass, and $m_{H(A)}$ are the masses of the CP-even (odd) scalars from the inert doublet, while $m_{H^{\pm}}$ is the charged scalar mass. Either of the neutral scalars can be the DM candidate in this IDM framework since DM observations can not probe the CP-behaviour. For the present analysis, we consider  the CP-even scalar $H$ as the DM candidate, which corresponds to negative values of $\lambda_5$ parameter. We define $\lambda_3+\lambda_4+\lambda_5 = \lambda_L$, which 
can be either positive or negative. The relations between the $\lambda$'s and the scalar masses get modified when the QED corrections are considered for both 
the scalar masses and scalar potential parameters. As the inert scalars do not couple to the SM quarks, higher-order QCD corrections are negligible for these parameters. 

Compared to the SM, only the scalar sector is modified in the IDM. Similar to the SM, $\lambda_1$ and $\mu_1^2$ are determined by $m_h\approx 125$ GeV and
$v\approx 246$ GeV. There are five free parameters in the scalar sector of the IDM viz. $\lambda_L$,  $\lambda_2$,  $m_A$,  $m_{H^{\pm}}$,  and $m_{H}$  that 
are expressed in terms of the five scalar potential parameters, $\mu_2^2$ and $\lambda_{2,3,4,5}$, as shown in Eq.~\ref{mass_relation}. 
The new doublet, being inert to the SM fermionic sector, does not introduce any additional new parameters in this set-up. 
In IDM the contribution of the self coupling parameter $\lambda_2$ is mostly limited to fixing unitarity and stability of the potential. It does not affect the scalar masses and their phenomenology. 
The Higgs portal coupling $\lambda_L$ to the chosen DM candidate $H$ determines the rate of the DM annihilation through the Higgs and therefore, 
is an essential parameter in the DM sector along with the DM mass $m_{DM} = m_H$. The collider phenomenology of the IDM depends on the scalar masses $m_{H^{\pm}}, m_A$ and $m_{H}$, as the mass differences between them play a significant role in proposing the search channels for different scenarios. 

\section{Constraints on the Inert Doublet Model}
\label{cons-idm}

The IDM parameter space is constrained from various theoretical as well as experimental considerations. 
In the model, we have an extra doublet which brings extra parameters in the scalar potential. Therefore, it is 
imperative to check whether the extended potential is bounded from below, {\emph i.e.}, stable at tree level 
along with the potential parameters being within the unitary and perturbative regime. With the presence of the extra doublet, 
oblique parameters should be re-examined with respect to the presence of a light DM and custodial symmetry breaking, respectively. The presence of light scalars can also upset the LEP constraints and the Higgs invisible decay limits. 
Since DM is present in the model, we should satisfy the observed relic density keeping
the DM-nucleon interactions below the DM direct detection reach.
	
%%%%%%%%%%%%%%%%%%%%%%%%%%%%%%%%%%%%%%%%%%%%%%%%%%%%%%

\subsection{Theoretical constraints}
\label{cons1} 
%%%%%%%%%%%%%%%%%%%%%%%%%%%%%%%%%%%%%%%%%%%%%%%%%%%%%%
%%%%%%%%%%%%%%%%%%%%%%%%%%%%%%%%%%%%%%%%%%%%%%%%%%%%%%
The scalar sector is modified in the IDM. We ensure the enlarged potential is stable, {\emph i.e.}, not unbounded from below 
and the global minimum is a neutral one. We also checked if the potential parameters are perturbative at the 
tree level along with satisfying unitarity bounds.  
%%%%%%%%%%%%%%%%%%%%%%%%%%%%%%%%%%%%%%%%%%%%%%%%%%%%%%

\vspace{0.2cm}
\noindent
\underline{\emph{Bounded from below:}}
%%%%%%%%%%%%%%%%%%%%%%%%%%%%%%%%%%%%%%%%%%%%%%%%%%%%%%
Theoretical constraints on quartic potential parameters ($\lm$'s) can arise from restricting the scalar potential in 
Eq.~\ref{v2hdm} not to produce large negative numbers for large field values, {\emph i.e.} $V>0$ $\forall$ $\Phi_i\to\pm\infty$. 
The mixed quartic terms can be combined to form complete square terms and demanding their coefficients to be positive,
leads to the following conditions\footnote{Alternately, for a scalar potential with many quartic couplings, one can consider formulating the copositive matrices to guarantee the boundedness of the potential~\cite{Chakrabortty:2013mha}.}.

\begin{eqnarray}
\lambda_1 > 0, \ \ \lambda_2 >0, \ \ \lambda_3 + 2 \sqrt{\lambda_1 \lambda_2} > 0, \ \ \lambda_3 + \lambda_4 +\lambda_5 + 2 \sqrt{\lambda_1 \lambda_2} >0 .    
\end{eqnarray}

Because of the introduction of new scalars, there are possibilities of having multiple minima. 
For the inert vacuum to be the global minimum, we restrict it from being charged 
by ensuring the condition
\begin{eqnarray}
\lambda_4 + \lambda_5 < 0 .   
\end{eqnarray}

 We also ensure that the global minimum is the inert vacuum as opposed to an inert-like one, with the imposition of the condition~\cite {Ginzburg:2010wa,Swiezewska:2012ej}
\begin{eqnarray}
\frac{\mu_1^2}{\sqrt{\lambda_1}} - \frac{\mu_2^2}{\sqrt{\lambda_2}} > 0 .   
\end{eqnarray}

%%%%%%%%%%%%%%%%%%%%%%%%%%%%%%%%%%%%%%%%%%%%%%%%%%%%%%
\vspace{0.2cm}
\noindent
\underline{\emph{Perturbativity and unitarity:}} 
We form the $S$-matrix with the amplitudes computed from the $2 \to 2$ scalar scattering processes taking into account all the other quartic terms in the scalar potential. 
The eigenvalues of the $S$-matrix turn out to be some combinations of these couplings. 
The perturbative unitarity constraints on those eigenvalues are $|\Lambda_i| \le 8 \pi$,
where the scattering matrix provides us the combinations as
\begin{align}   
\Lambda_{1,2} &= \lambda_3 \pm \lambda_4; \hspace{0.4cm}
\Lambda_{3,4} = \lambda_3 \pm \lambda_5; \hspace{0.4cm}
\Lambda_{5,6} = \lambda_3 + 2 \lambda_4 \pm 3 \lambda_5; \nn\\
\Lambda_{7,8} &= - \lambda_1 -\lambda_2 \pm \sqrt{(\lambda_1 - \lambda_2)^2 + \lambda_4^2}; \nn\\  
\Lambda_{9,10} &= - 3 \lambda_1 - 3 \lambda_2 \pm \sqrt{9(\lambda_1 - \lambda_2)^2 + (2 \lambda_3 +\lambda_4)^2}; \nn\\
\Lambda_{11,12} &= - \lambda_1 - \lambda_2 \pm \sqrt{(\lambda_1 - \lambda_2)^2 + \lambda_5^2} .
\end{align}

%%%%%%%%%%%%%%%%%%%%%%%%%%%%%%%%%%%%%%%%%%%%%%%%%%%%%%
%%%%%%%%%%%%%%%%%%%%%%%%%%%%%%%%%%%%%%%%%%%%%%%%%%%%%%

\subsection{Collider constraints}
\label{sec:colcons}
%%%%%%%%%%%%%%%%%%%%%%%%%%%%%%%%%%%%%%%%%%%%%%%%%%%%%%
Precision measurements at the LEP and the LHC contributed in pinning down the trace of new physics effects in different forms. 
The effects of hierarchical heavy BSM scalar mass spectrum and the presence of a lighter DM are under consideration. 
After the discovery of the Higgs boson, the LHC also measured its properties. Two such measurements, 
the Higgs decay to $\gamma\gamma$ and its invisible decay are important to consider in the context of IDM. 

\vspace{0.2cm}
\noindent
\underline{\emph{Oblique correction constraints:}}
The oblique parameters $S, T$ and $U$, proposed by Peskin and Takeuchi~\cite{Peskin:1991sw}, are different combinations of the oblique corrections, {\emph i.e.}, radiative corrections to the two-point functions of the SM gauge bosons. The $S$ parameter encodes the running of the neutral gauge boson two-point functions ($\Pi_{ZZ}, \Pi_{Z\gamma},\Pi_{\gamma \gamma}$) in the lower energy range, from zero momentum to the $Z$-pole. Therefore, the $S$ parameter is sensitive to the presence of light particles 
with masses below $m_Z$, which is the case here due to the presence of the light DM. The $T$ parameter, on the other hand,
measures the difference between the $WW$ and the $ZZ$ two-point functions, $\Pi_{WW}$ and $\Pi_{ZZ}$, at zero momentum. Mass splitting of the scalars inside a $\mathrm{SU}(2)_L$ doublet breaks the custodial symmetry which modifies $T$. 
In the IDM, the mass splittings between the neutral and the charged scalars are controlled by the $T$ parameter.
The experimentally measured values of oblique parameters that we use in our analysis are~\cite{PDG18}:
\begin{align}
S = 0.05 \pm 0.11;\hspace{0.4cm} T = 0.09 \pm 0.13; \hspace{0.4cm} U = 0.01 \pm 0.11. 
\end{align}

\vspace{0.2cm}
\noindent
\underline{\emph{$h\to\gamma\gamma$ signal strength constraint:}} 
The signal strength for the $h\to\gamma \gamma$ channel is given by the following ratio \cite{Arhrib:2012ia,Swiezewska:2012eh},
\begin{align}
   R_{\gamma\gamma} = \frac{\sigma(pp \to h)}{\sigma(pp \to h_{\textrm{SM}})} \times \frac{\textrm{BR}(h \to \gamma \gamma)}{\textrm{BR}(h_{\textrm{SM}} \to \gamma \gamma)} .
\end{align}
In the IDM, the Higgs production rate is similar to that of the SM, as it is gluon fusion dominated in both the models. So, 
in the IDM, the ratio can be approximated as
\begin{align}
   R_{\gamma\gamma} = \frac{\textrm{BR}(h \to \gamma \gamma)_{\rm IDM}}{\textrm{BR}(h \to \gamma \gamma)_{\rm SM}} 
\end{align}
Combined CMS and ATLAS fit in the diphoton channel provides a $2\sigma$ limit on this observable as~\cite{Khachatryan:2016vau},
\begin{align}
   R_{\gamma\gamma} = 1.14^{+0.38}_{-0.36}.
\end{align}
Presence of a charged Higgs in the $h\to\gm\gm$ decay loop can induce a significant shift in this ratio for large values of 
$hH^+H^-$ coupling. In the IDM, this coupling depends on $\lambda_3$ which is also related to the charged Higgs mass and 
this parameter is constrained from the allowed range of the ratio $R_{\gamma \gamma}$ that can deviate from unity. \footnote{ In the IDM, only the Higgs decay rate to $\gamma \gamma $ can deviate from the SM value at the leading order. As that deviation is within the experimental limit, the Higgs boson here easily satisfies all Higgs signal data.}

\vspace{0.2cm}
\noindent
\underline{\emph{Constraint from the Higgs invisible decay:}} 
Another constraint from the Higgs data, applicable for the scenario when Higgs can decay to a pair of DM particles with a mass  $m_{DM} < m_h/2$. The invisible decay width is given by
\begin{equation}
\Gamma (h \rightarrow \text{Invisible})= {\lambda^2_L v^2\over 64 \pi m_h} 
\left(1-\frac{4\,m^2_{DM}}{m^2_h} \right)^{\frac{1}{2}}.
\end{equation}
The latest ATLAS constraint on the invisible Higgs decay is \cite{Aad:2015gba}
$$\text{BR} (h \rightarrow \text{Invisible}) = \frac{\Gamma (h \rightarrow \text{Invisible})}{\Gamma (h \rightarrow \text{Invisible}) + 
	\Gamma (h \rightarrow \text{SM})} < 22 \%.$$
In the case for light DM when the Higgs decay to a pair of DM particles is kinematically allowed, this limit can
significantly constrain the parameter space of the IDM.
%%%%%%%%%%%%%%%%%%%%%%%%%%%%%%%

\vspace{0.2cm}
\noindent
\underline{\emph{LEP bounds:}} 
A reinterpretation of the neutralino search results at the LEP-II has ruled out the parameter regions \cite{Lundstrom:2008ai, Belanger:2015kga} that satisfy the following three conditions
\begin{equation}
m_H < 80 \rm \ GeV, \ m_A < 100  \ \rm GeV \ and \ m_A - m_H > 8 \ \rm GeV.
\label{eq:lep}
\end{equation} 
Reinterpretation of chargino search results at the LEP-II has put a bound \cite{Pierce:2007ut} on the charged Higgs mass as,
\begin{equation}
m_{H^{+}} > 70 \rm \ GeV.
\end{equation} 
A hierarchical IDM scalar spectrum is not restricted from these constraints. Moreover, due to large mass gap in the spectrum, $Z \to H A, W^{\pm} \to H H^{\pm}, W^{\pm} \to A H^{\pm} $ off-shell decays have a negligible effect on the total width of the $W$ and $Z$ bosons, that are very precisely measured at the LEP experiments.

%%%%%%%%%%%%%%%%%%%%%%%%%%%%%%% 
%%%%%%%%%%%%%%%%%%%%%%%%%%%%%%% 
%%%%%%%%%%%%%%%%%%%%%%%%%%%%%%% 
\subsection{Astrophysical constraints}

%%%%%%%%%%%%%%
This model contains a DM candidate, the CP-even scalar in $\Phi_2$. Therefore, astrophysical constraints on this model consist of the DM relic density and the direct probe of DM in the Xenon and LUX experiments.

\vspace{0.2cm}
\noindent
\underline{\emph{Relic density:}}
There are unputdownable observational pieces of evidence of the presence of DM in a vast range of length scale, starting from intergalactic rotation curve to the latest Planck experiment data. That suggests the current density of the DM comprises approximately $26\%$ energy budget of the present Universe. 
The observed abundance of DM is usually represented in terms of density parameter $\Omega$ as \cite{Ade:2015xua} 
\begin{equation}
\Omega_{\text{DM}} \mathfrak{h}^2 = 0.1187 \pm 0.0017
\label{dm_relic}
\end{equation}
where the observed Hubble constant is $\mathcal{H}_0 = 100 \, \mathfrak{h} \, km \, s^{-1} \, Mpc^{-1}$. The rate of DM annihilation to the SM particles is inversely proportional to the relic of the DM, and therefore constraints are imposed to avoid the overproduction of the relic in the IDM. We compute the DM relic density numerically with MicrOmega~\cite{Belanger:2001fz}, implementing the IDM details there.

%%%%%%%%%%%%%%

\vspace{0.2cm}
\noindent
\underline{\emph{Direct detection constraints:}}
Along with the constraints from the relic abundance measurement in the Planck experiment, there exist strict bounds on the 
DM-nucleon cross section from the DM direct detection experiments like Xenon100 (Xenon1T)~\cite{Aprile:2012nq} and more recently from LUX~\cite{Akerib:2016vxi}. For scalar DM considered in this work, the spin independent DM-nucleon scattering cross section mediated by the SM Higgs is given as \cite{Barbieri:2006dq}
\begin{equation}
 \sigma_{\text{SI}} = \frac{\lambda^2_L f^2}{4\pi}\frac{\mu^2 m^2_n}{m^4_h m^2_{DM}},
\label{sigma_dd}
\end{equation}
where $\mu = m_n m_{DM}/(m_n+m_{DM})$ is the DM-nucleon reduced mass and $\lambda_L=(\lambda_3+\lambda_4+\lambda_5)$ is the quartic coupling involved in the DM-Higgs interaction. A recent estimate of the Higgs-nucleon coupling is $f = 0.32$ \cite{Giedt:2009mr}, although the full range of allowed values is $f=0.26-0.63$ \cite{Mambrini:2011ik}. As shown in Fig.~\ref{fig:DM} later, the Xenon1T upper bound on the DM-nucleon scattering can put a stringent limit on the allowed $\lambda_L$ values that constrain the Higgs-DM coupling. 

%%%%%%%%%%%%%%%%%%
\begin{figure}[!t]
	\begin{center}
		\subfigure[]{\includegraphics[height=6cm,width=8cm]{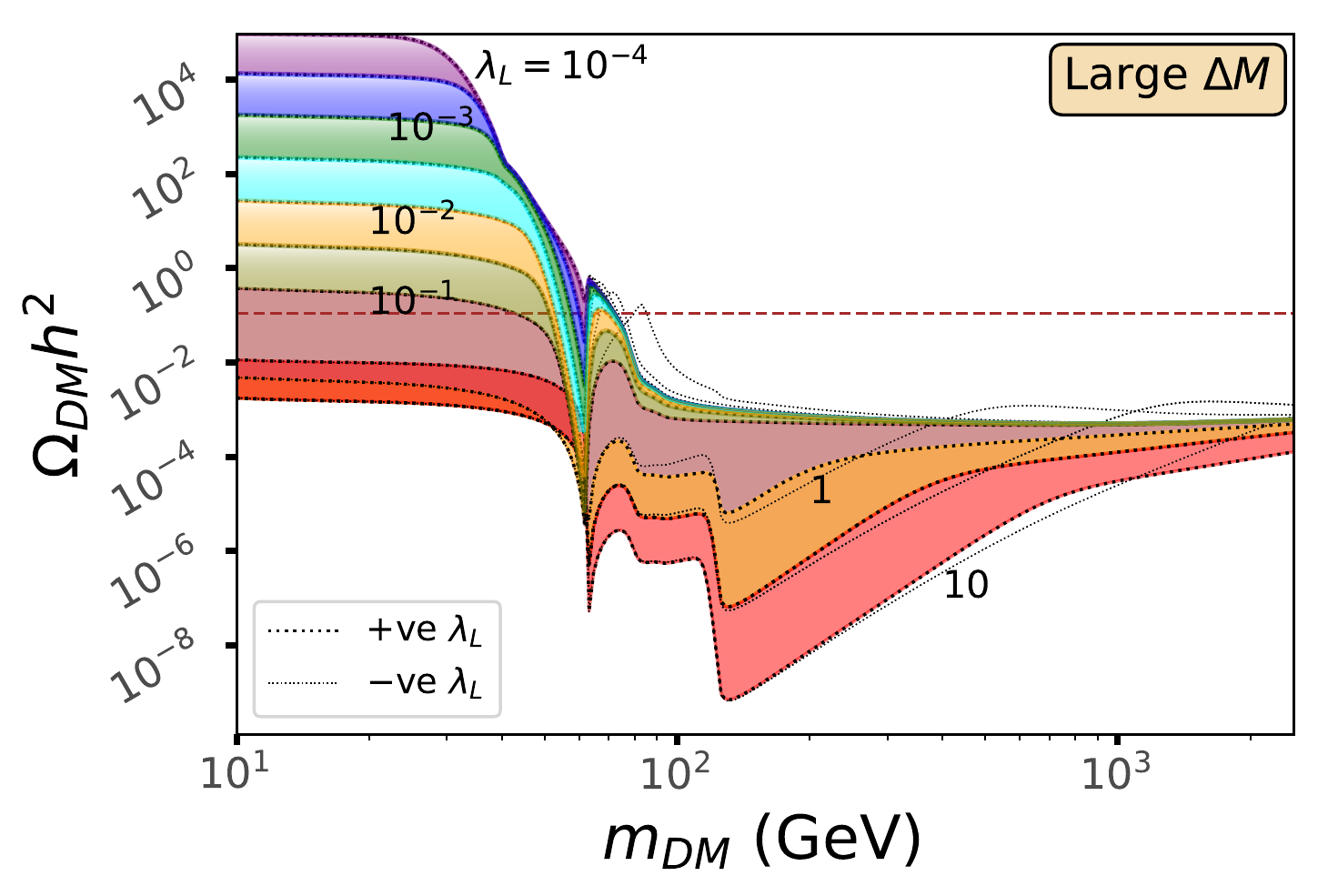}\label{fig:FD2DM}}\hspace{0.4cm}
		\subfigure[]{\includegraphics[height=6cm,width=8cm]{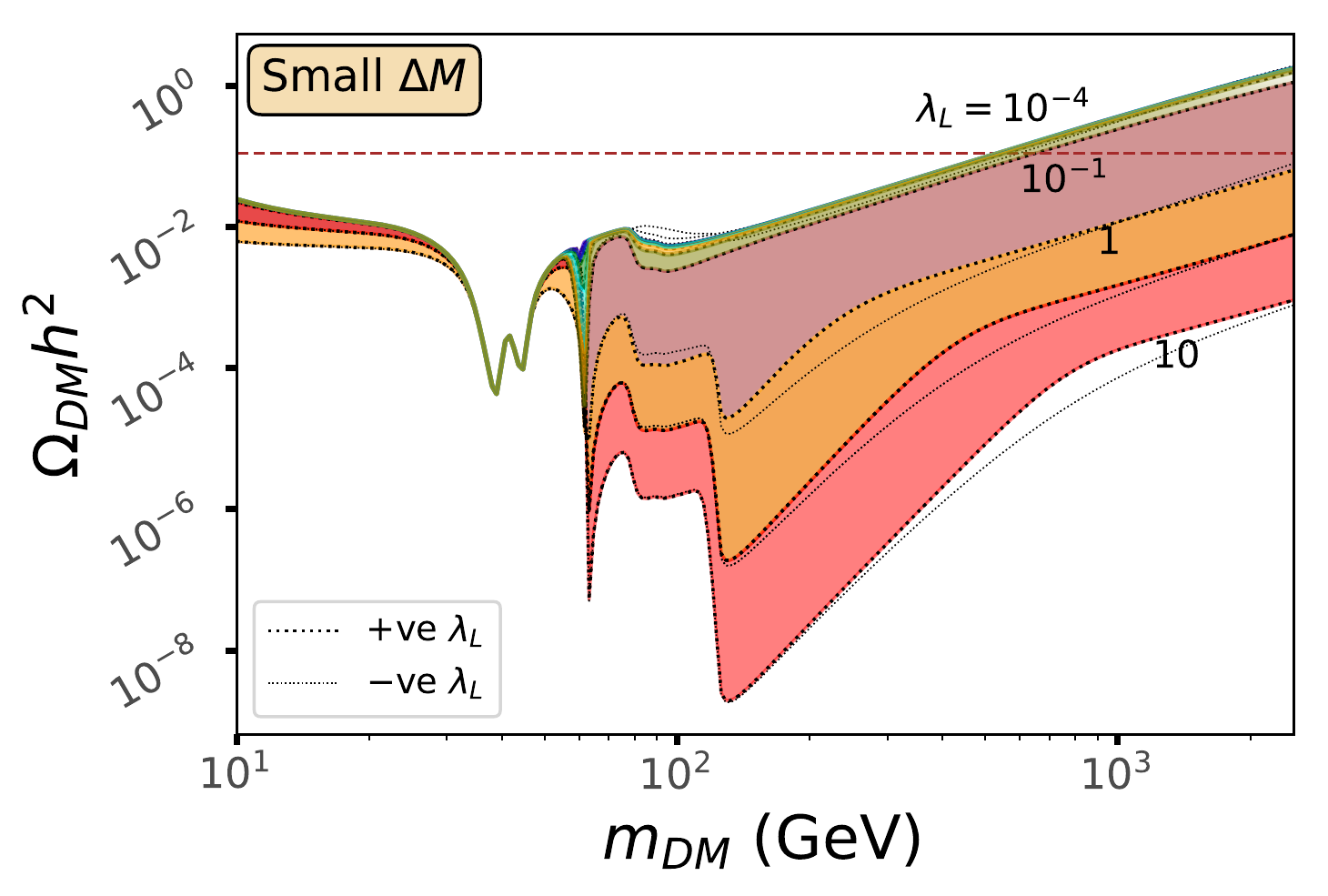}\label{fig:FD1DM}}
	\end{center}
	\caption{Variation of relic density $\Omega_{DM} h^2$ shown as a function of dark matter mass $m_{DM}$ in the inert doublet model. Band of colours with thick dotted lines considering different $\lambda_L$ values. Corresponding negative values of $\lambda_L$ are shown with thin dotted lines.  DM relic abundance strongly depends upon the mass difference  $\Delta M$ between dark matter candidates from additional BSM scalar masses. Left and  right plots correspond to the large and small values of it, respectively.  One can identify four DM paradigms inside the IDM parameter space, as discussed in Table~\ref{table:Param}.}
	\label{fig:DM_possible_cases}
\end{figure}
%%%%%%%%%%%%%%%%%%

%%%%%%%%%%%%%%%%%%%%%%%%%%%%%%%%%%%%%%%
\section{Possible searches and benchmarking}
\label{sec:BP}
 We explore the DM paradigm inside the IDM, discussing the variation of DM relic density with various model parameters. The relic density dependence on DM mass for both the hierarchical and degenerate nature of the BSM mass spectrum is presented in Fig.~\ref{fig:DM_possible_cases} for different $\lambda_L$ values. The nature of the mass spectrum is quantified by two mass differences $\Delta M_1 \equiv m_{H^{\pm}} - m_{DM} $ and $\Delta M_2 \equiv m_A - m_{DM}$, which are assumed to be equal ($\Delta M$) in the plots. Effects of the hierarchical mass spectrum with $\Delta M = 100~$GeV and the degenerate mass spectrum with $\Delta M = 1~$GeV on DM relic density are depicted in Fig.~\ref{fig:FD2DM} and Fig.~\ref{fig:FD1DM}, respectively.
We also point out how sign-reversal of $\lambda_L$ can alter the relic density dependence on the DM mass. 

%%%%%%%%%%%%%%%%%%%%%%%%%%%%%%%%%%%%%%%%%%%%%%%%%%%%%%%%%%%%%%%%%%%%%%%%%%%%%%%%%%%%%%%%%%%%%%%%%%%%%%%%%%%%%%%%%%%%%%%%%%%%%
 We roughly divide the DM paradigm of the IDM into four different cases depending on the DM mass and the nature of the mass spectrum, specified by $\Delta M_1, \Delta M_2$. These four cases are showcased in Table~\ref{table:Param}.
In each scenario, we discuss the thermal DM relic abundance along with the phenomenological study done to probe the BSM scalars. 

%%%%%%%%%%%%%%%%%%%%%%%%%%%%%%%%%%%%%%%%%%%%%%%%%%%%%%%%%%%%%%%%%%%%%%%%%%%%%%%%%%%%%%%%%%%%%%%%%%%%%%%%%%%%%%%%%%%%%%%%%%%%%
\vspace{0.2cm}
\noindent
\underline{\emph{Case~I:}}  
We first consider a case of light DM with mass, $m_{\textrm{DM}} \lesssim 80$~GeV together with all other heavy scalars within a narrow mass range. This case is severely constrained from the LEP data which rules out $m_{\textrm{DM}} < 45$~GeV. 
Precise LEP measurements of the $Z$-width constrains $Z \to A H$ decay together with the conditions in Eq.~\ref{eq:lep} .
Even for the DM mass above $45$~GeV, the degenerate nature of the spectrum ensures that all the inert scalars take part in the 
co-annihilation processes and reduce the relic density to somewhat below $10\%$ of the total relic. 
Instead of both of the mass differences $\Delta M_i$ tiny, if one of them is taken to be large, only the scalar with smaller $\Delta M$ dominantly affect the extent of DM co-annihilation. 
Sharp dip appears when the DM mass is at $m_h/2$ due to the resonant production peak in the DM annihilation through the Higgs portal. Furthermore, some additional  shallow dips in the relic density are also observed when the $WW$ and the $ZZ$ annihilation modes 
open up, enhancing the annihilation cross section. 
In this low mass region, the DM annihilation is contributed dominantly through the Higgs portal, and thus the sign of $\lambda_L$ does not affect the relic density.
This scenario is explored at the LHC in the mono-jet signal, as discussed in Ref.~\cite{Belyaev:2018ext}.

%%%%%%%%%%%%%%%%%%
\begin{table}[!t]
\centering
\begin{tabular}{|c||c|c|c|}
\hline
\backslashbox{~~$\Delta M$}{$M_{DM}$~~}
& Small & Large  \\[0ex]
\hline\hline 
      &${\fbox{\bf Case~I}}~~ M_{DM} ~ <$    80 GeV & ${\fbox{\bf Case~III}}~~M_{DM} ~ \sim$    550 GeV 
\\  Small ~  & $~~~~~~~~~\Delta  M ~ \sim {\cal O} (1-10)$ GeV  & $~~~~~~~~~\Delta  M \sim {\cal O} (1)$ GeV
\\    & $~~~~~~~~Relic ~Density \sim 10\%$  &  $~~~~~~~~Relic~ Density \sim 100\%$ 

%\\    & $\pk{Collider~searches}$  &  $\pk{Collider~searches}$ 
%\vspace{0.1cm}
\\
  
\hline
      &${\fbox{\bf Case~II}}~~M_{DM} ~ <$    80 GeV  &  ${\fbox{\bf Case~IV}}~~M_{DM} ~ \sim$ 550 GeV  
\\  Large ~  & $~~~~~~~~~\Delta  M \sim {\cal O} (100) $ GeV  & $~~~~~~~~~\Delta  M \sim {\cal O} (10-100) $ GeV
\\    & $~~~~~~~~Relic~ Density \sim 100\%$  &  $Relic~ Density \sim 1\%$ 
%\\    & $\pk{Collider~searches}$  &  $\pk{Collider~searches}$
 \\
\hline
\end{tabular}
\caption{Illustration of four DM paradigms inside the IDM parameter space, comparing DM mass as well as scalar mass hierarchy. Available DM density as a fraction of the required relic density is also pointed out for these cases. We study the phenomenologically interesting but challenging region marked by ${\bf Case~II}$.}
\label{table:Param}
\end{table}
%%%%%%%%%%%%%%%%%%

\vspace{0.2cm}
\noindent
\underline{\emph{Case-II:}} 
 Here, we consider the light DM with the hierarchical scalar mass spectrum, {\emph i.e.} large mass differences ($\Delta M_i$) with both of the other heavy scalars. Because of this large mass difference between $H$ and $A$/$H^\pm$, the LEP $Z$-width measurements do not constrain such a low 
DM mass. DM annihilates only through the Higgs portal and therefore for tiny $\lambda_L$, the relic is overproduced. However, the entire relic density can be described at larger $\lambda_L$ values, which are progressively bounded from the DM direct detection data from LUX and Xenon1T. Contrasting this with the degenerate case as pointed out in `Case-I', here the co-annihilation effect is absent in the annihilation cross section and increases the relic density to produce a full relic in the range
$m_{\textrm{DM}}\sim 53-70$~GeV depending on different $\lambda_L$ values. Phenomenologically this parameter range is quite interesting although detection of such very light DM along with much heavier other scalars is challenging at the collider. One has to encounter a very small production cross section along with an extremely large SM background where the signal characteristics are very background-like. The LHC potential of this case is studied in the dijet plus MET channel in Ref.~\cite{Poulose:2016lvz}. Here, we take up this scenario for further analysis.

\vspace{0.2cm}
\noindent
\underline{\emph{Case-III:}} 
If we move towards the heavier DM regime, a degenerate mass spectrum can provide full relic density at around $m_{\textrm{DM}} \sim 550$~GeV \footnote{Recently, this limit is brought down to the DM mass $\sim 400$~GeV, as shown in Ref.~\cite{Borah:2017dfn}, by introducing right handed neutrinos, whose late decay to the DM compensates the under-produced DM relic density seen previously.}. Exact mass depends strongly on the value of $\lambda_L$ parameter. From a $10\%$ relic for $m_{\textrm{DM}} \sim 100$~GeV, it steadily increases as the $HH \to WW, ZZ$ annihilations open up and the cross section decreases with mass. 
 The $HHVV$ coupling turns out to be $\lambda_{HHV_i V_i} \sim (4m_{\textrm{DM}}\Delta M_i/v^2 + \lambda_L)$ in the limit DM and other heavy scalars are mass degenerate, {\emph i.e.}, $\Delta M_1 \approx \Delta M_2 \to 0$. We explored this part of the parameter space earlier with both $\Delta M_i = 1~$GeV. Even if the DM annihilation rate increases with the DM mass, that increase is strongly suppressed due to tiny mass 
differences between the different BSM scalars in a nearly degenerate mass spectrum. The DM relic density, along with being inversely dependent on the annihilation cross section, also is directly proportional to the DM mass. 
Therefore, the interplay of these two competing effects finally ends up in a gradual increase in the DM relic density. The quartic coupling essentially depends only on $\lambda_L$ in $\Delta M_i \to 0$, even then a $\lambda_L$ sign reversal does not affect the DM pair annihilation. This scenario is phenomenologically interesting but quite challenging to probe at the LHC.
This extremely compressed scenario can be probed at the LHC with identifying the charged track signal of a long-lived charged scalar~\cite{CMS:2014gxa}.

\vspace{0.2cm}
\noindent
\underline{\emph{Case-IV:}} 
 In the heavier DM regime with hierarchical mass spectrum where both the mass differences are large, {\emph e.g.}, $\Delta M_1 \approx \Delta M_2 \sim 100~$GeV, the annihilation cross section increases with the DM mass. This happens due to rapid increase of the DM-gauge boson quartic couplings with its mass, {\emph i.e.}, $\lambda_{HHV_i V_i} = 2(2 m_{\rm DM} + \Delta M_i) \Delta M_i/v^2 + \lambda_L$, for any general $\Delta M_i$, which is a result of  the large mass difference between the BSM scalars. 
 Enhancement in the DM annihilation leads to drop of the relic density with increasing $m_{\rm DM}$ producing up to a few percents of the full observed value. Here, $\lambda_L$ dependence is mostly overshadowed by large mass differences and does not affect much. 
 In the case of very distinct choices of $\Delta M_i$ values, the DM annihilation would be dominated by the scalar having a larger mass difference through this enhanced coupling while the other one would contribute negligibly. Therefore the DM scenario in Case-III can be envisaged as a limiting case of Case-IV.

Among the four DM scenarios in the IDM as described above and also summarized in Table~\ref{table:Param}, two cases have emerged as phenomenologically exciting. Light DM ($m_{\rm DM}\sim 50-80~$GeV) with hierarchical mass spectrum with 
a substantial mass gap ($\Delta M_i \gtrsim 100$ GeV) with other heavy scalars can provide the full observed DM relic density. 
On the other side, we get a rather heavy DM ($m_{DM}\sim 550~$GeV) with an extremely degenerate mass spectrum, which can also provide the required relic density. Both the scenarios are challenging to probe, as the heavier BSM scalars are difficult to produce in the inert model and essentially confront with large irreducible SM backgrounds. 

\begin{figure*}[h!]
	\centering
	\includegraphics[scale=0.35]{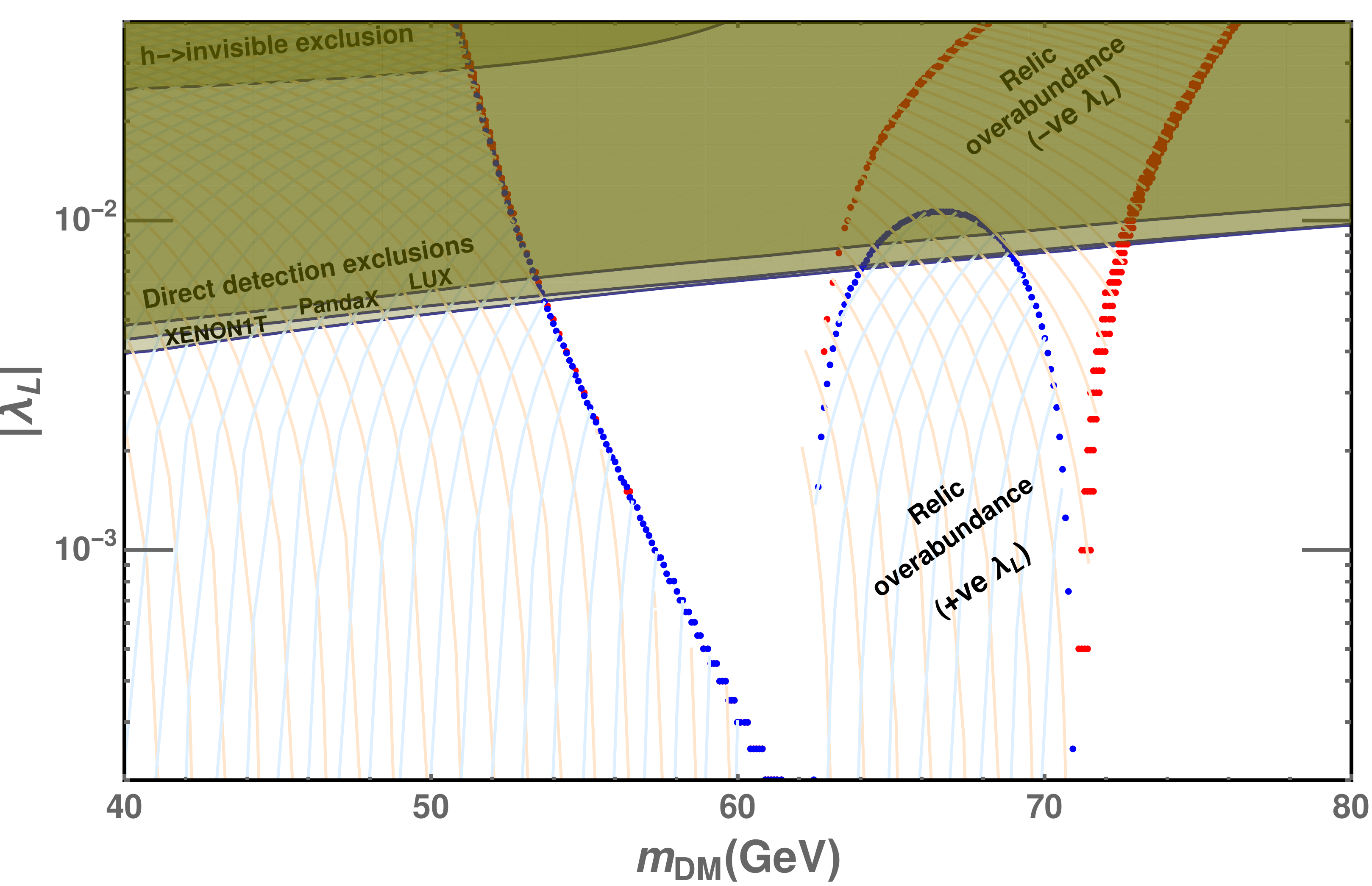}
	\caption{Allowed DM relic abundance in $m_{DM} - |\lambda_L|$ parameter space extracted for a set of $M_H^{\pm}, M_A$ and $\lambda_2$ values. Blue (red) dots are the points where observed abundance is exactly satisfied as in Eq.~\ref{dm_relic} for +ve (-ve) values of $\lambda_L$. The shaded area under this curve represents over-abundance and is thus excluded. Two other shaded regions at the upper portion of the plot are excluded from invisible decay of Higgs and DM direct detection constraints from XENON1T, Panda, and LUX, respectively.}
	\label{fig:DM}
\end{figure*}

Now, we particularly focus on the low DM mass (50-70 GeV)  with hierarchical mass spectrum {\emph i.e.}, $\Delta M_1, \Delta M_2 \gtrsim 200$~GeV for our phenomenological study. 
To demonstrate the exact numerical evaluation, in Fig.~\ref{fig:DM}, we explore the $m_{\rm DM}-\lambda_{L}$ parameter plane of the IDM for a light DM with $\Delta M_i = 100$~GeV applying the constraints from the DM relic density measurements, the DM direct detection experiments and the constraint from the Higgs invisible decay. This choice of $100$~GeV is representative since major annihilation modes for DM are through the Higgs portal and the parameter space of this plot is equally valid for larger $\Delta M$ choices. Blue (red) dots are the points where the observed DM relic abundance is exactly satisfied as in Eq.~\ref{dm_relic} for +ve (-ve) values of $\lambda_L$. The shaded area under this curve represents DM over-abundance and thus is excluded. Two other constraints can come from the invisible decay of the Higgs and the DM direct detection constraints from XENON1T, which are shown in the same plane in two other shaded regions in the upper portion of the plot, respectively. The DM direct detection constraints from LUX (Xenon1T) put stringent upper bound on $\lambda_L$, for all values of light DM.  All other constraints described above, provide weaker bounds in this parameter space \cite{Krawczyk:2013jta}.

%%%%%%%%%%%%%%%%%%%%%%%%%%%%%
%%%%%%%%%%%%%%%%%%%%%%%%%

%%%%%%%%%%%%%%%%%%%%%%%%% scale=0.715
\begin{figure*}[t!]
	\begin{center}
		\includegraphics[scale=0.45]{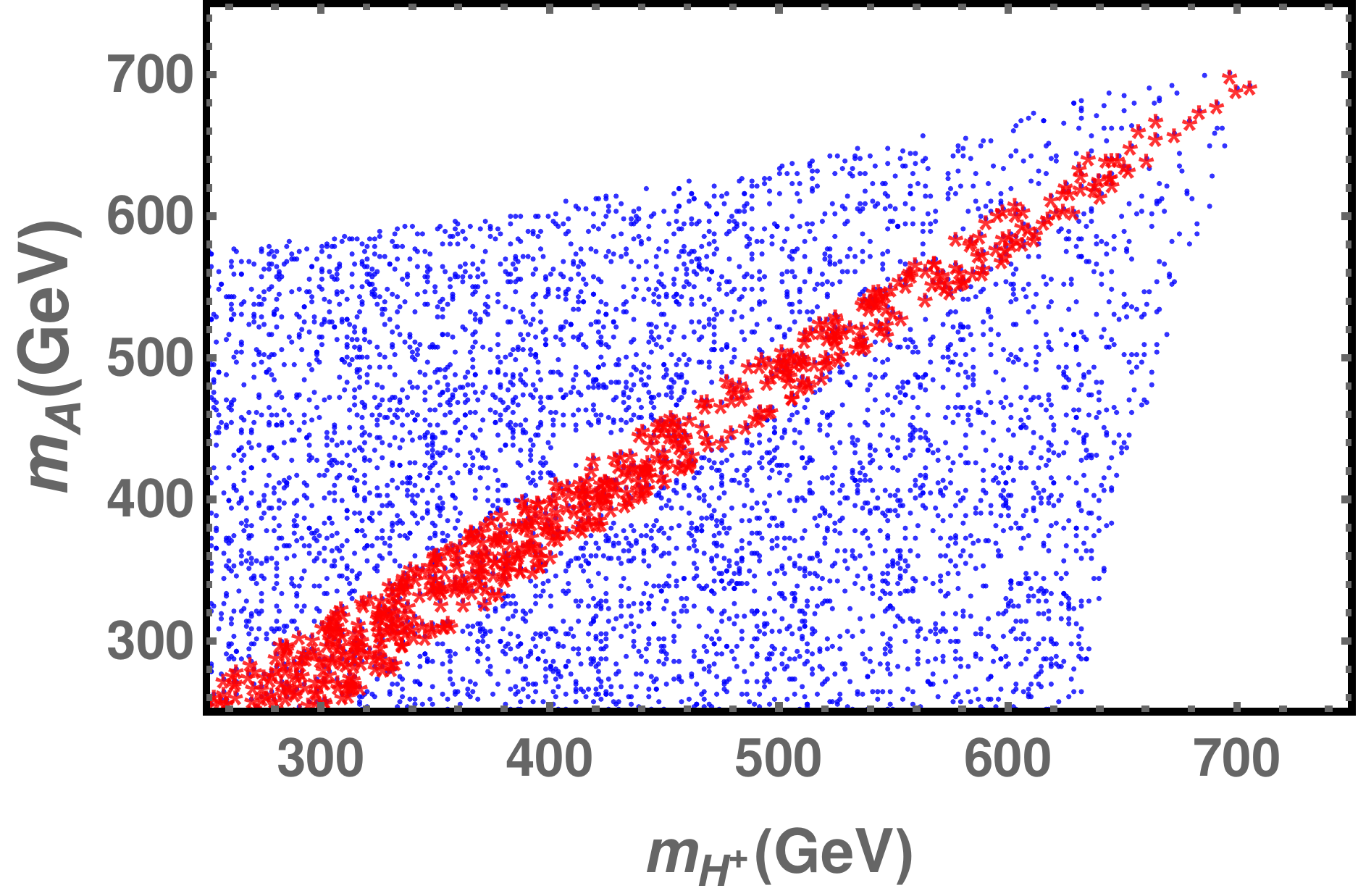}
		
		\caption{ The blue scatter plot shows the limits from positivity and perturbativity constraints in the $(M_H^{\pm}, M_A)$ plane after fixing the DM mass and $\lambda_L$ for all benchmark points. The red dots represent the allowed parameter space after imposing the constraints from the oblique parameters $S$, and $T$. Similar allowed parameter space is found for other benchmark points.}
		\label{fig:Unitarity_ST}
	\end{center}
\end{figure*}
%%%%%%%%%%%%%%%%%%%%%%%%%

With our understanding of allowed DM mass and $\lambda_L$ parameter in the light DM scenario, we now attempt to comprehend other remaining parameters. 
To do so, we set these parameters to a particular choice from the allowed region of the relic density plot %$M_H = 53.71 $ and $\lambda_L = 5.4 \times10^{-3}$
and then perform a scan over the remaining three parameters comprising of heavy scalar masses ($M_H^{\pm}, M_A$) and $\lambda_2$. One such frame of the allowed parameter space after imposing the theoretical constraints (unitarity, perturbativity etc.) along with the $R_{\gamma\gamma}$ constraint from Secs.~\ref{cons1},~\ref{sec:colcons} are shown by the blue scatter plots in Fig.~\ref{fig:Unitarity_ST}. The red dots in the same plot represent the values of $M_H^{\pm}$ and $M_A$, which satisfy the oblique parameter constraints. 
These constraints, primarily through the $T$ parameter, force these heavy scalar masses $M_H^{\pm}$ and $M_A$ to be almost degenerate.
  
To study the specific low mass DM scenario within the IDM, we choose a set of seven benchmark points (BPs) from the allowed parameter space. These BPs covering heavy scalar mass between 250 to 550 GeV along with the corresponding input DM mass and $\lambda$ parameter are summarized in Table~\ref{tab:BP}. 
It is worth noting that the choice of $M_H$ and $\lambda_L$ is for the theoretical and experimental consistency, but the collider analysis proposed in this paper holds equally good for all the allowed points in Fig.~\ref{fig:DM}. Radiative correction to the DM-Higgs portal coupling is calculated in Ref. \cite{Banerjee:2016vrp}, which allows slightly more parameter space.

%%%%%%%%%%%%%%%%%%%%%%%%%
\begin{table}[t!]
 \centering
 \renewcommand{\arraystretch}{1.5}
\begin{tabular}{|c|c|c|c|c|c|c|c|}
	\hline
	Parameters  & BP1  & BP2 & BP3 & BP4 & BP5  & BP6 & BP7\\
	\hline
	\hline
$M_{H^{\pm}}$(GeV)&  255.3 &  304.8   &  350.3 & 395.8  &  446.9 & 503.3 & 551.8 \\ 

		\hline
$M_A$(GeV)     &  253.9	& 302.9    & 347.4& 395.1  &  442.4 & 500.7& 549.63 \\

	 	 \hline
$\lambda_2$ &  1.27 &  1.07 & 0.135 & 0.106 &  3.10 & 0.693  & 0.285 \\
	 	 	
	 	 	 \hline
\end{tabular}
 \caption{Input parameters $\lambda$ and the relevant scalar masses for some of the chosen 
 benchmark points satisfying all the constraints coming from  DM, Higgs, theoretical calculations and low energy experimental data as discussed in the text. 
  All the mass parameters are written in units of GeV. Standard choice of the other two parameters are fixed  at $M_H = 53.71$ GeV and $\lambda_L = 5.4 \times10^{-3}$.
  }
 \label{tab:BP}
\end{table}
%%%%%%%%%%%%%%%%%%%%%%%%%%%%%%%%%%%%%%%
\section{Collider analysis}
\label{coll}
%%%%%%%%%%%%%%%%%%%%%%%%%%%%%%%%%%%%%%

We make use of various publicly available HEP packages for our subsequent collider study aiming for a consistent, reliable  detector level analysis. We implement the IDM Lagrangian in {\tt FeynRules}~\cite{Alloul:2013bka} to create the UFO~\cite{Degrande:2011ua}
model files for the event generator {\tt MadGraph5 (v2.5.5)}~\cite{Alwall:2014hca} which is used to generate all signal and background events. These events are generated at the leading order (LO) and the 
higher-order corrections are included by multiplying appropriate QCD $K$-factors. We use {\tt CTEQ6L1}~\cite{Pumplin:2002vw} parton distribution 
functions for event generation by setting default dynamical renormalization and factorization scales used in
{\tt MadGraph5 }\cite{madgraph_scale}. Events are passed through {\tt Pythia8}~\cite{Sjostrand:2006za} to perform showering and 
hadronization and matched up to two to four additional jets for different processes using {\tt MLM} matching 
scheme~\cite{Mangano:2006rw,Hoche:2006ph} with virtuality-ordered Pythia showers to remove the double counting 
of the matrix element partons with parton showers. The matching parameter, $\texttt{QCUT}$ is appropriately determined 
for different processes as discussed in Ref. \cite{madgraph_matching}. Detector effects are simulated using 
{\tt Delphes}~\cite{deFavereau:2013fsa} with the default CMS card.
Fatjets are reconstructed using the {\tt FastJet}~\cite{Cacciari:2011ma} package by clustering {\tt Delphes} 
tower objects. We employ Cambridge-Achen (CA)~\cite{Dokshitzer:1997in} algorithm with radius parameter $R = 0.8$ 
for jet clustering. Each fatjet is required to have $P_T$ at least 180 GeV.
We use the adaptive Boosted Decision
Tree (BDT) algorithm in the {\tt TMVA} framework~\cite{Hocker:2007ht} for MVA.

\subsection{Signal topology}
\label{ssec:sigtop}

\begin{figure}[t!]
\begin{center}
\subfigure[]{\includegraphics[height=4cm,width=6cm]{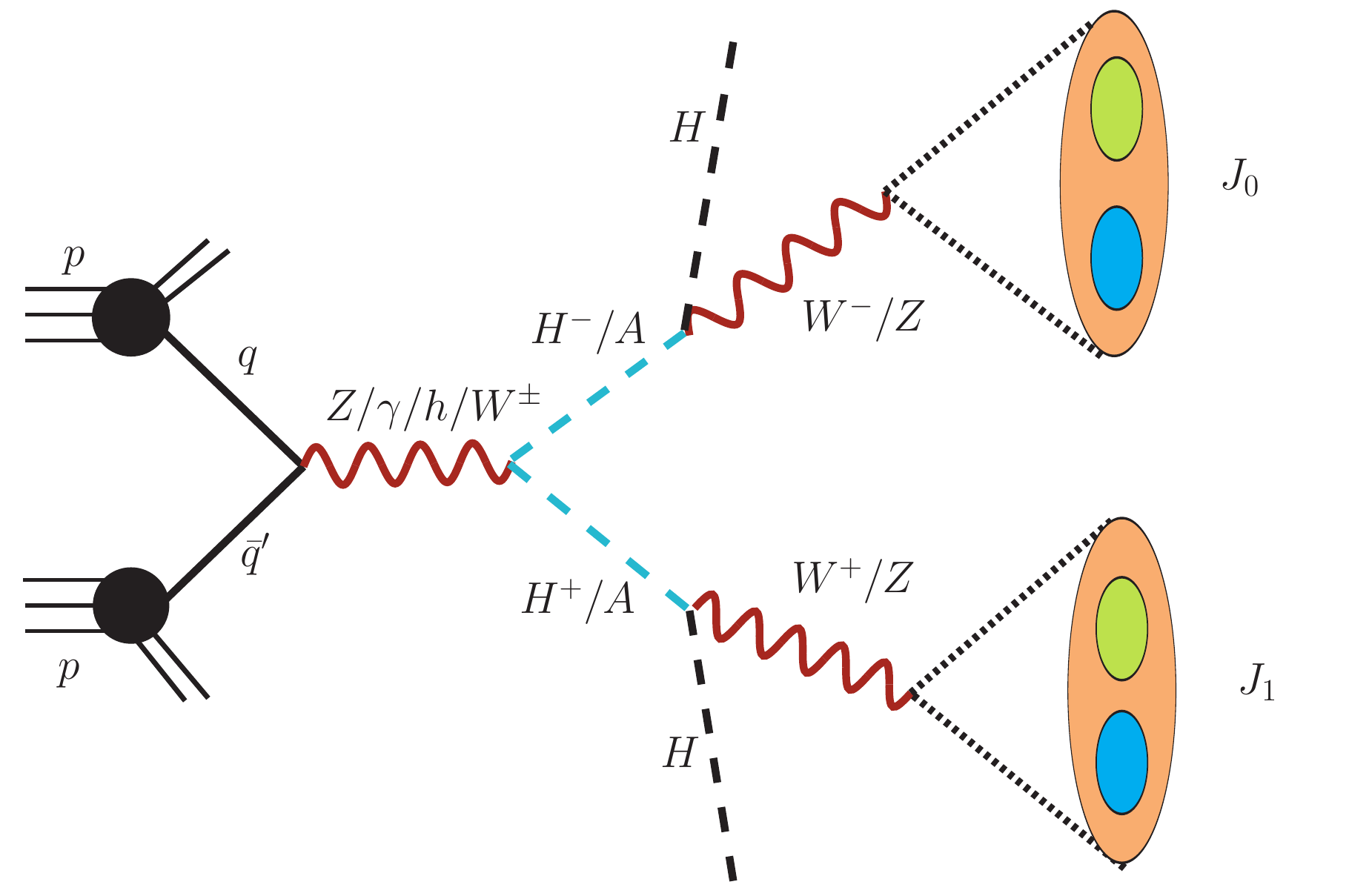}\label{fig:FD2FJ}}\hspace{0.4cm}
\subfigure[]{\includegraphics[height=4cm,width=6cm]{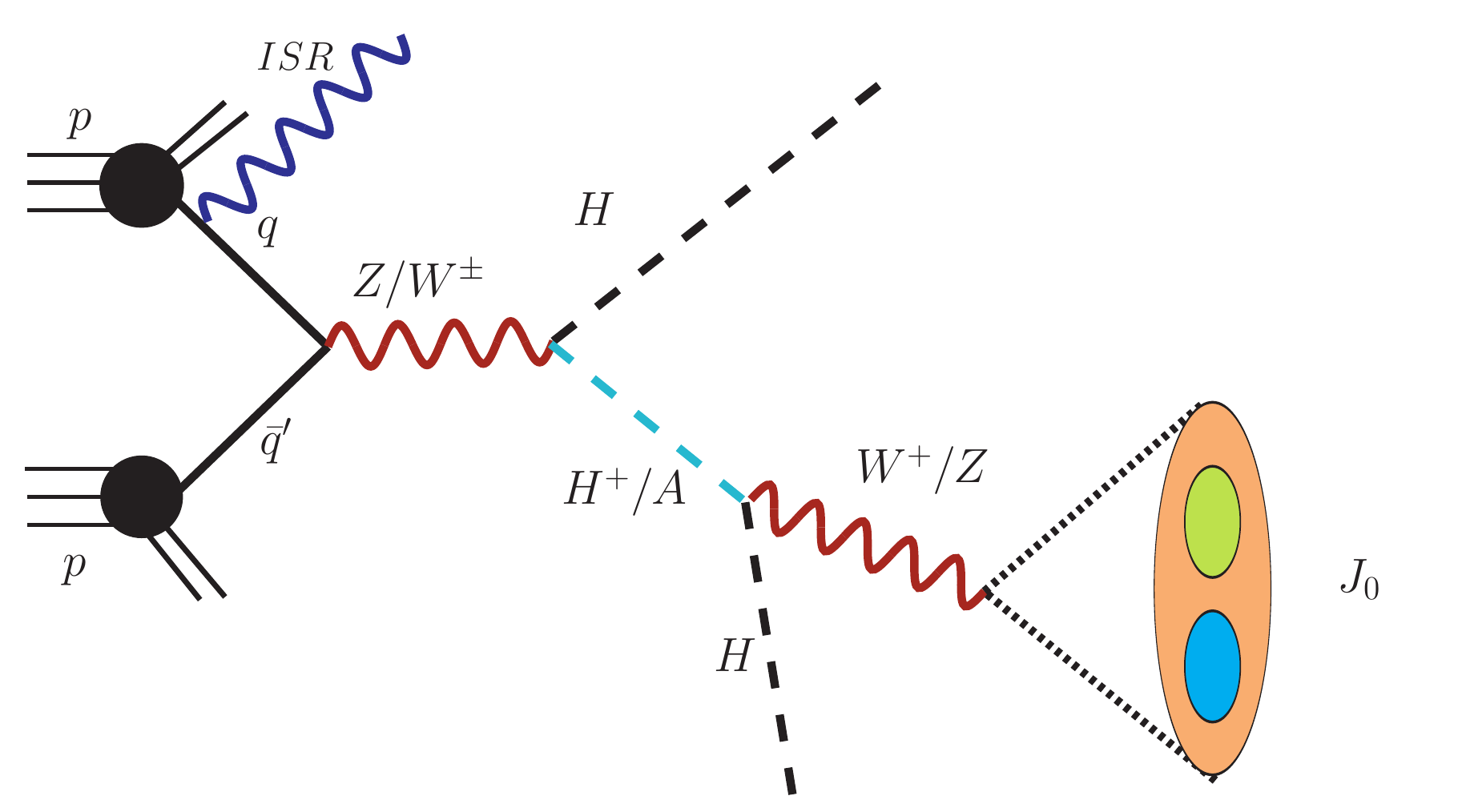}\label{fig:FD1FJ}}
\end{center}
\caption{Representative parton level diagrams of (a) di-$V$-jet plus missing-energy ($2 J_V+\slashed{E}_T$) and (b) 
mono-$V$-jet plus missing-energy ($1 J_V+\slashed{E}_T$).}
\label{fig:FD}
\end{figure}

As discussed in the introduction, the hierarchical mass pattern in the IDM scalar sector, ({\emph i.e.}, $M_A\sim M_{H^{\pm}}\gg M_H$)
provides us with interesting final states. Once a pair of heavy scalars (or one heavy scalar associated with DM candidate) are  produced at the LHC, they eventually decay dominantly producing two (or one) boosted vector bosons, each of which is decaying hadronically and thus producing $V$-jet ($J_V$) where $V=\{W,Z\}$. These boosted $V$-jets are always associated with 
large MET ($\slashed{E}_T$), an outcome of our inability to detect the DM pair at the detector. Representative
Feynman diagrams of these signal topologies are demonstrated in Fig.~\ref{fig:FD}. Among them, it must already be clear 
to the readers 
that the $1 J_V+\slashed{E}_T$ channel alone, although being cross-section-wise bigger than $2 J_V+\slashed{E}_T$, has
less sensitivity at the LHC due to overwhelmingly large SM background. Therefore, we primarily focus on the  
$2 J_V+\slashed{E}_T$ channel where the large background can be tamed down by employing jet substructure variables
in an MVA framework. Our signal is not pure $1 J_V+\slashed{E}_T$ or $2 J_V+\slashed{E}_T$ topologies rather it is an admixture of
both processes. Note that the baseline selections (defined in Sec. \ref{ssec:selec}) are designed keeping $2 J_V+\slashed{E}_T$ topology
in mind. This keeps a large fraction of events from the $1 J_V+\slashed{E}_T$ topology. In doing this we gain in the signal, but at the same time one can avoid extremely large background related to the $1 J_V+\slashed{E}_T$ topology, ({\emph i.e.}, by demanding at least one $J_V$ instead of two). Before moving on to the actual analysis, we give some useful details about these two signal
topologies. 
%%%%%%%%%%%%%%%%%%
%%%%%%%%%%%%%%%%%%

\vspace{0.2cm}
\noindent
\underline{$2 J_V+\slashed{E}_T$ channel:}
This final state can arise in the IDM for the aforementioned benchmarks from the following three different channels,
\begin{align}
pp &\to AH^{\pm} \to (ZH)(W^{\pm}H)\equiv 2 J_V+\slashed{E}_T\nn\\
pp &\to H^{+}H^{-} \to (W^{+}H)(W^{-}H)\equiv 2 J_V+\slashed{E}_T\\
pp &\to AA \to (ZH)(ZH)\equiv 2 J_V+\slashed{E}_T\nn\ .
\end{align}  
Here, $A$ and $H^{\pm}$ decay to $ZH$ and $W^\pm H$, respectively. 
As $Z$ and $W$ are originating from a heavy resonance, it is possible that they have sufficient boost to be reconstructed 
in a large radius jet. We do not distinguish a $Z$-jet or a $W$-jet and call them $V$-jet  as we always select fatjets
with a broad mass range. A $V$-jet possesses a two prong substructure, {\emph i.e.}, energy distribution is centered around two subjet axes. 
We utilize the $N$-subjettiness ratio $\tau_{21}$ (defined later) to tag $V$-jets.

\vspace{0.2cm}
\noindent
\underline{$1 J_V+\slashed{E}_T$ channel:}
This final state can arise from the following two different channels,
\begin{align}
pp &\to H^{\pm}H \to (W^{\pm}H)H\equiv 1 J_V+\slashed{E}_T\nn\\
pp &\to AH \to (ZH)H\equiv 1 J_V+\slashed{E}_T\ .
\end{align} 

Extra jets can arise in the final state due to initial state radiation (ISR) and can form another fatjet. So these channels can potentially 
mimic the $2 J_V+\slashed{E}_T$ final state. We generate matched samples of 
this signal with up to two additional jets in the final state. 
In this topology, only one of the two fatjets has the $V$-jet like structure and the other jet originates 
from the QCD radiation which mimics the fatjet characteristics. 
We find that the contributions to the $2 J_V+\slashed{E}_T$ final state from the $1 J_V+\slashed{E}_T$ topologies are 
quite significant and sometimes bigger than the $2 J_V+\slashed{E}_T$ contribution itself after our final selection.
This is mainly due to bigger production cross-sections of $pp\to AH,H^{\pm}H$ processes and the tail events which satisfy the 
fatjet criteria of our analysis~\footnote{The motivation to choose $2 J_V+\slashed{E}_T$ channel is that one has larger  features than in the case of  $1 J_V+\slashed{E}_T$ to handle the enormous background, where $1 J_V+\slashed{E}_T$ also contributes to the signal $2 J_V+\slashed{E}_T$ when the extra QCD jet mimics as a fatjet. The $1 J_V+\slashed{E}_T$ is explored in the searches in Refs. \cite{Khachatryan:2016mdm,Aaboud:2018xdl}.}.
	
The leading order production cross sections for the signal processes discussed above for different BPs 
are given in Table~\ref{tab:BPxsecn}. We have used NLO QCD $k$-factors of 1.27 and 1.50 for the $q\bar{q}$ and the $gg$ initiated productions for the signal~\cite{Hespel:2014sla}. 
\begin {table}[t!]
\centering
\renewcommand{\arraystretch}{1.1}

\begin{tabular}{|c||c|c||c|c|c|}
	\hline
	Benchmark   &\multicolumn{5}{c|}{{$\sg(pp\to xy)$ (fb)}} \\ \cline{2-6}
	Points & $ AH^{0}$ & $ H^{\pm}H^{0}$              & $ AH^{\pm}$ & $ H^{+}H^{-}$ & AA  \\ \hline
	BP1	  & 34.54    &   62.53   & 12.62    & 7.96   & 0.50     \\ \hline	
	BP2    & 18.71    & 34.12    &  6.22 & 4.23 & 0.40 \\ \hline
	BP3	  & 11.43    & 20.84   & 3.50 & 2.59 & 0.34 \\ \hline
	BP4	  &  7.11    & 13.32   & 2.05 & 1.70 & 0.28 \\ \hline
	BP5	  &  4.63    & 8.44    & 1.22    & 1.13    & 0.24      \\ \hline		
	BP6	  & 2.84    & 5.32   & 0.71 & 0.76 & 0.19 \\ \hline
	BP7	  &  1.95  &  3.70   & 0.45 & 0.56 & 0.16 \\ \hline		
	\hline
\end{tabular}
\caption{Production cross sections for the signal processes that contribute to the $1 J_V+\slashed{E}_T$ and $2 J_V+\slashed{E}_T$ final states at the 14 TeV LHC. 
These numbers are for $pp\to xy$ level before the decay of IDM scalars.}
\label{tab:BPxsecn}
\end{table}	
% % % % % % %%%%%%%%%%%%%%%%%%%%%%%%%%%%%%%%%%%%%%%%%%%%%%%%%%%%5

\subsection{Backgrounds}
\label{subsec:bg}

For our hybrid signal discussed in the introduction as well as in Sec. \ref{ssec:sigtop}, major backgrounds come from the following SM processes which we discuss briefly below. All these backgrounds are carefully included in our analysis.

\vspace{0.2cm}
\noindent
\underline{$V + jets$:}\\
There are the following two types of mono-vector boson processes that contribute dominantly in the background. 

\begin{itemize}

\item
\underline{$Z + jets$:}
This is the most dominant background in our case. We generate the event samples by simulating the inclusive $pp\rightarrow  Z + jets\to \nu\nu+jets$ process matched up to four extra partons.
Here, invisible decay of $Z$ gives rise to a large amount of $\slashed{E}_T$ and QCD jets mimic as fatjets. 

\item 
\underline{$W + jets$:}
This process also contributes significantly in the background when $W$ decays leptonically, and the lepton does not satisfy the selection criteria. 
This is often known as the lost lepton background. The neutrino comes from the $W$-decay and contributes to missing energy and QCD jets mimic as fatjets.
We generate the event samples by simulating inclusive $pp\rightarrow  W + jets\to \ell_{(e,\mu)} \nu+jets$ process matched up to four extra partons.
\end{itemize}

In order to get statistically significant background events coming from the tail phase space region with large $\slashed{E}_T$, we apply a hard cut of $\slashed{E}_T > 100$~GeV at the generation level to generate these background events.

\vspace{0.2cm}
\noindent
\underline{$VV + jets$:}\\
Different diboson processes like $WZ$, $WW$ and $ZZ$ also mimic the signal and contribute to the SM background. The $pp\to WZ$ process contributes most 
significantly among 
these three diboson channels when $W$ decays hadronically, and $Z$ decays invisibly. We call this background as $W_hZ_\nu$. Similarly, $W_hW_\ell$, where one
$W$ decays hadronically and the other leptonically, and $Z_hZ_\ell$ (a hadronic $Z$ and a leptonic $Z$) can also contribute to the SM backgrounds when leptons remain
unidentified. All the diboson processes are generated up to two extra jets with {\tt MLM} matching. 
In this case, one of the fatjets can come from the hadronic decay of $V$, and the other can come from the hard partons. 

\vspace{0.2cm}
\noindent
\underline{Single top:}\\
Single top production in the SM includes three types of processes viz. top associated with $W$ ({\emph i.e.} $pp\to tW$ process), 
$s$-channel single top process ({\emph i.e.}, $pp\to tb$) and $t$-channel single top process ({\emph i.e.}, $pp\to tj$). 
Among these, the associated production $tW$ contributes significantly in the SM background for our signal topologies. 

\vspace{0.2cm}
\noindent
\underline{$tt + jets$:}\\
This can be a background for our signal topologies when it decays semileptonically, {\emph i.e.}, one of the top decays  leptonically and the other decays hadronically. 
This background contains $b$-jets. We control this background by applying a $b$-veto. This background always has one $V$-jet. Another fatjet can 
originate from an untagged $b$-jet or QCD radiation. 

%%%%%%%%%%%%%%%%%%%%%%%%%

\begin{table}[t!]
\begin{center}
\begin{tabular}{|c|c|c|}
\hline
\multicolumn{2}{|c|}{Background process} & $\sg$ (pb)\\ \hline
\multirow{2}{*}{$V + jets$~\cite{Catani:2009sm,Balossini:2009sa}}  & $Z + jets$  &  $6.33 \times 10^4$ [NNLO] \\ \cline{2-3} 
                   & $W + jets$  & $1.95 \times 10^5$ [NLO] \\ \hline
\multirow{3}{*}{$VV + jets$~\cite{Campbell:2011bn}}  & $WW + jets$  & 124.31 [NLO]\\ \cline{2-3} 
                   & $WZ + jets$  & 51.82 [NLO]\\ \cline{2-3} 
                   & $ZZ + jets$  &  17.72 [NLO]\\ \hline
\multirow{3}{*}{Single top~\cite{Kidonakis:2015nna}}  & $tW$  &  83.1 [N$^2$LO] \\ \cline{2-3} 
                   & $tb$  & 248.0 [N$^2$LO]\\ \cline{2-3} 
                   & $tj$  & 12.35 [N$^2$LO]\\ \hline
Top pair~\cite{Muselli:2015kba}  & $tt + jets$  & 988.57 [N$^3$LO]\\ \hline
\end{tabular}
\caption{Cross sections for the background processes considered in this analysis at the 14 TeV LHC. These numbers are shown with the QCD correction order
provided in brackets.}
\label{tab:Backgrounds}
\end{center}
\end{table}

Apart from the above background processes, we also calculate the contributions from triboson and QCD multijet processes. However, these contributions are found to be insignificant as compared to the background discussed above, and therefore, we neglect the contribution of these backgrounds in the analysis. The production cross-sections with higher order QCD corrections for all the background processes considered in this analysis at the 14 TeV LHC are listed in Table~\ref{tab:Backgrounds}.

%%%%%%%%%%%%%%%%%%%%%%%%%%%%%%%%%%%%%%%%%%%%%%
\section {Cut-based analysis}
\label{sec:CBA}

We perform a CBA to estimate the sensitivity of observing the IDM signatures at the high luminosity LHC runs. 
It is evident that the signal cross sections are too
small compared to the vast SM background. Therefore, one needs sophisticated kinematic observables for the isolation of signal events from the background events.
Our signal processes always include at least a hadronically decaying vector boson that can provide a $V$-like fatjet. Therefore, we make use of the jet substructure 
variables for our purpose.

\subsection {$V$-jet tagging: jet substructure observables}

Jet substructure observables have emerged as a powerful technique to search for new physics signatures at colliders. 
In our case, boosted $W$ and $Z$ bosons, originated from the decay of heavy IDM scalars ($H^\pm$, $A$), give rise to 
collimated jets that can form a large radius jet (fatjet). These fatjets have two-prong substructures. 
We utilize two jet substructure observables viz. the jet-mass ($M_J$) and $N$-subjettiness ratio ($\tau_{21}$). 
The $M_J$ is a viable observable to classify the $V$-jets from the fatjets originated from QCD jets. We calculate the jet mass as $M_J = (\sum_{i  \in  J} P_i)^2$ where $P_i$ are the four-vector of energy hits in the calorimeter.
The discrimination power of $M_J$ reduces if extra contribution comes from the parton, which does not actually 
originate from the $V$-decay. This results in broadening of the peak in the $M_J$ distributions. Different jet grooming techniques are proposed to remove these softer and wide-angle radiations, such as  trimming, pruning, and filtering\cite{Krohn:2009th,Ellis:2009su,Ellis:2009me,Butterworth:2008iy}. We choose pruning for grooming the fatjets. 

\begin{figure*}[ht!]
 	\includegraphics[scale=0.38]{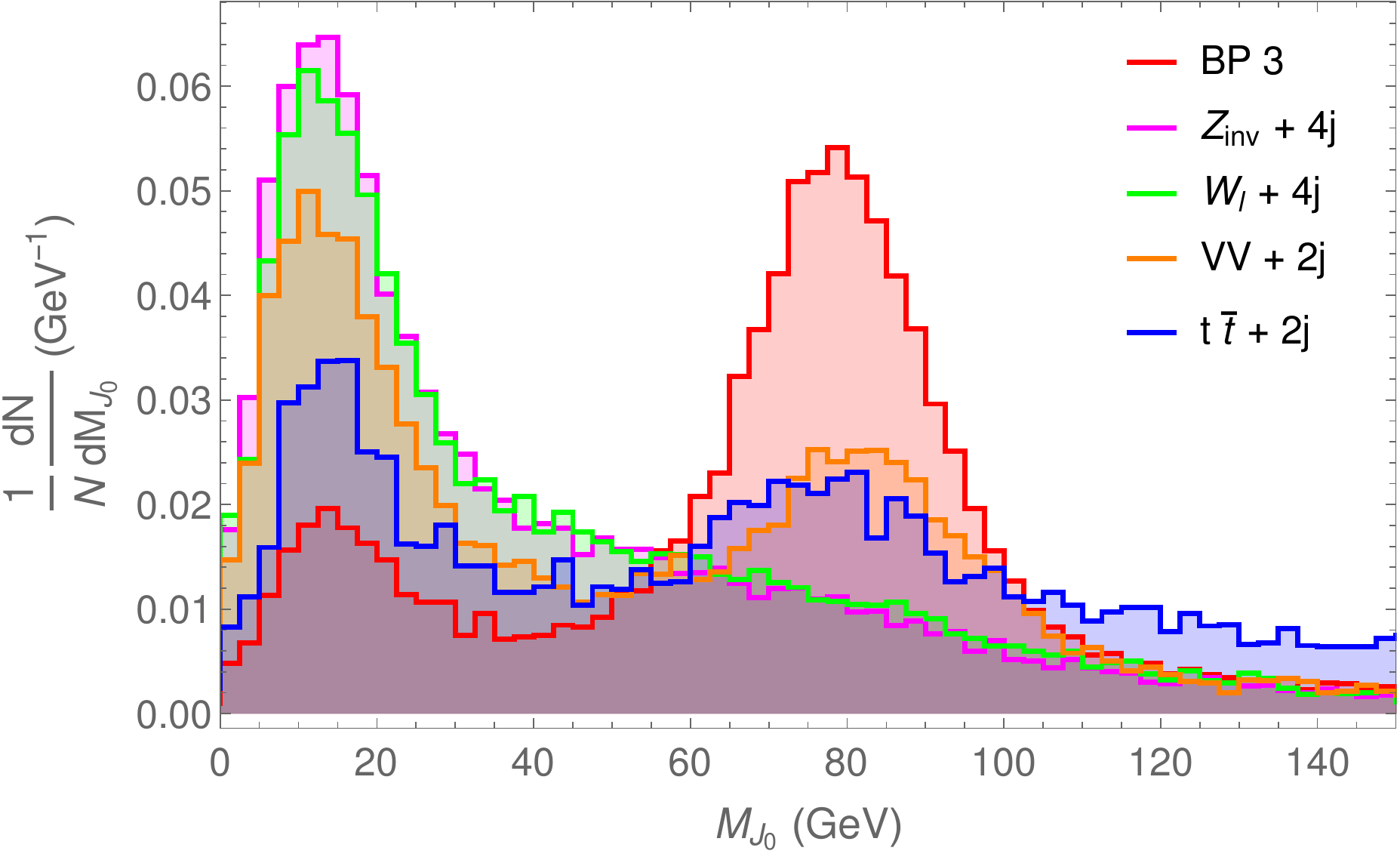}~~
 	\includegraphics[scale=0.38]{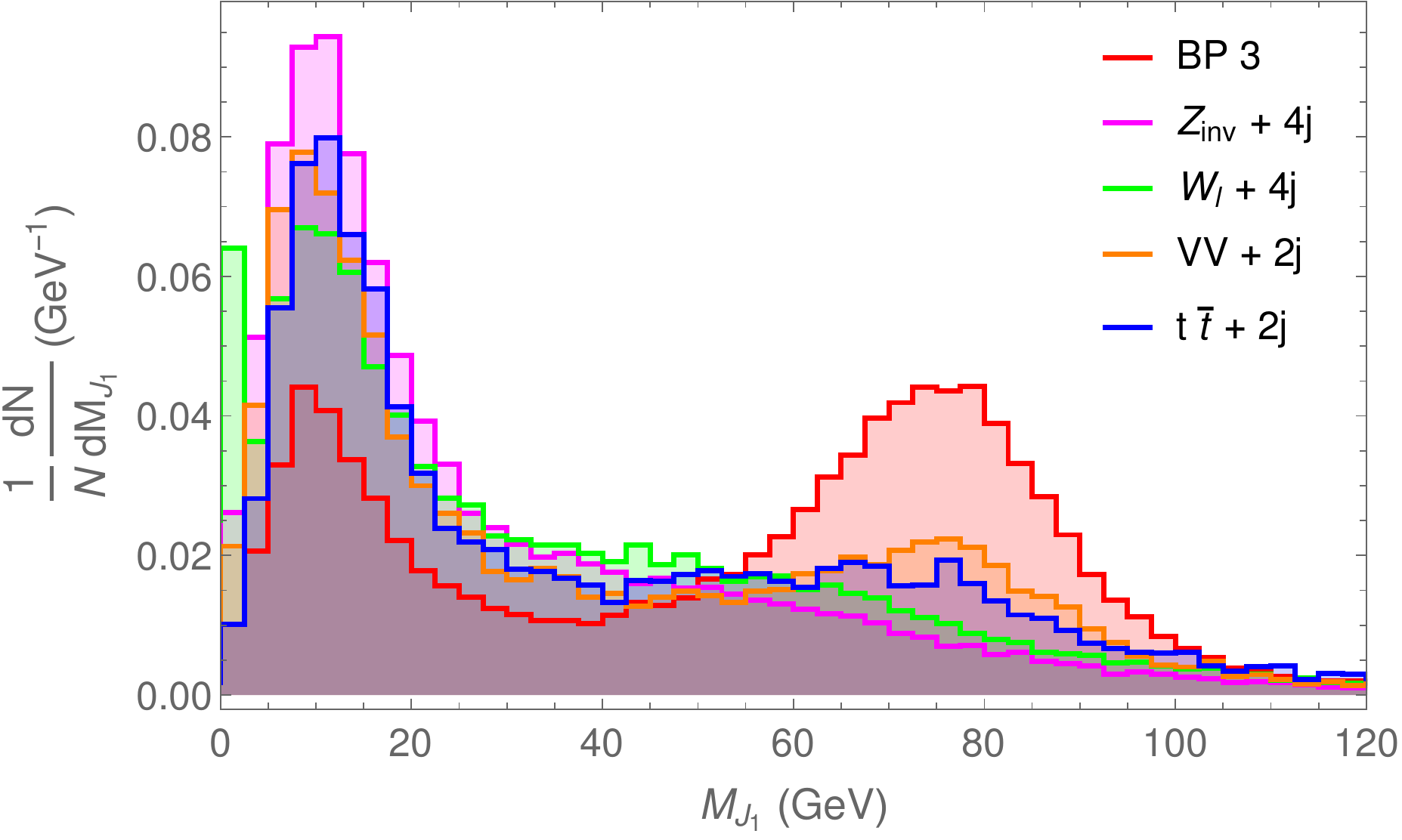}
 	\caption{Normalized distributions for invariant mass of leading fatjet $M_{J_0}$ (left) and and subleading fatjet $M_{J_1}$ (right) after the $baseline~selection~cuts$.}
 	\label{fig:dist_var1}
 \end{figure*}

\vspace{0.2cm}
\noindent
\underline{Pruned jet mass:}\\
We performed the pruning with the standard method as prescribed in Refs.~\cite{Ellis:2009su,Ellis:2009me} 
to clean the softer and wide-angle emission by rerunning the algorithm and vetoing on such recombinations. 
At each step of recombination, one calculates the two variable $z$ and $\Delta R_{ij}$, where $z$ is defined as 
$z = min(P_{Ti},P_{Tj})/P_{T_{i+j}}$ and $\Delta R_{ij}$ is the angular separation between two 
proto-jets. If $z < z_{cut}$ and $\Delta R_{ij} > R_{fact}$  then the $i$-th and $j$-th proto-jets are not recombined and the softer one is discarded. Here,  $z_{cut}$ and $R_{fact}$ are parameters of the pruning algorithm.  
We have taken the default values of $R_{fact}=0.5$ and $z_{cut}=0.1$ as suggested in Ref.~\cite{Ellis:2009su}. 
In Fig.~\ref{fig:dist_var1}, we show the distributions for pruned jet mass for signal (BP3) and the important backgrounds. It is evident from these distributions that the peak around $80-90$ GeV reflect the $V$-mass peak for the 
signal whereas for most of the background processes the peaks below 20 GeV reflect 
the fatjets mimic from a single prong hard QCD jet.

\begin{figure*}[ht!]
 	\includegraphics[scale=0.38]{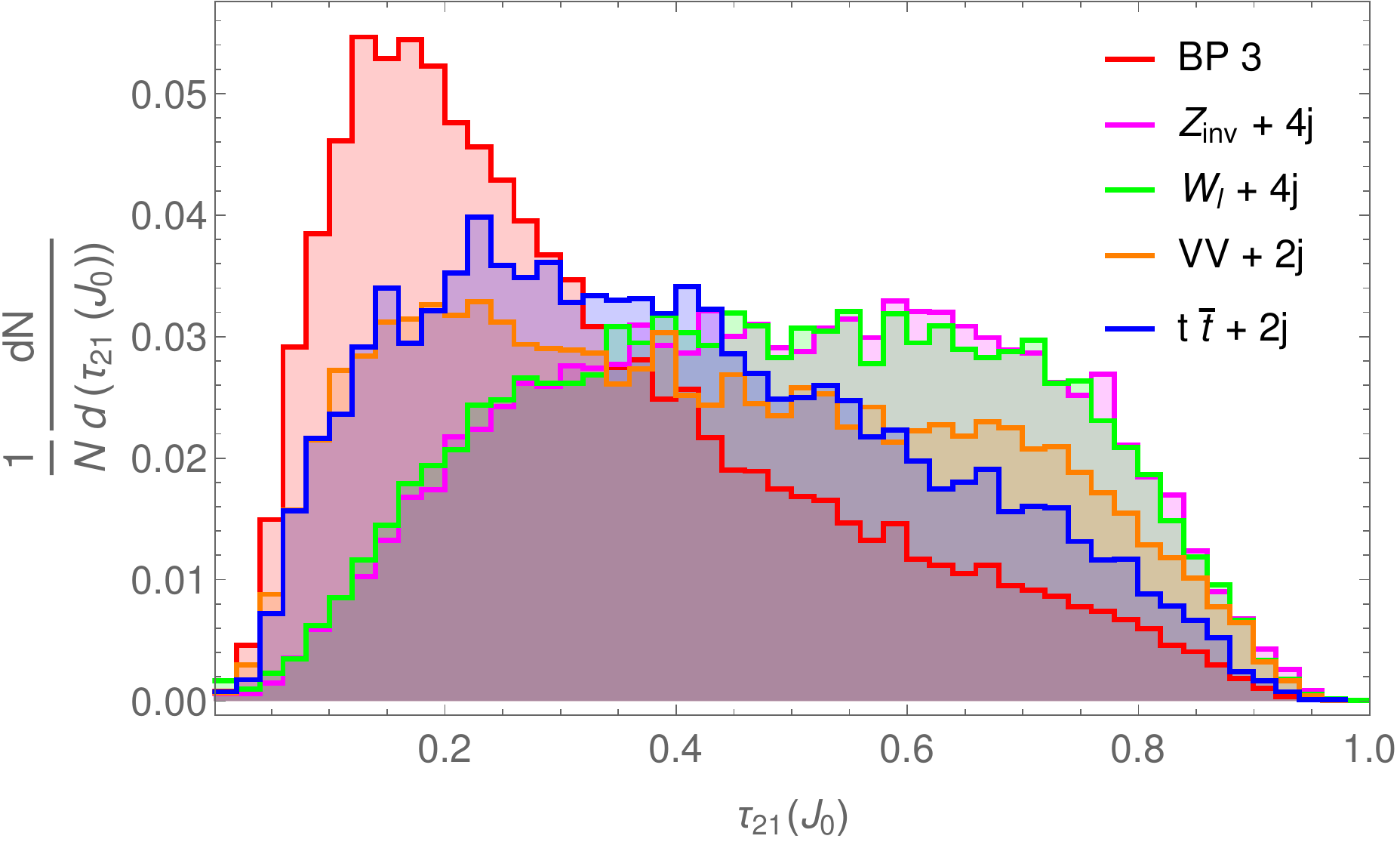}~~
 	\includegraphics[scale=0.38]{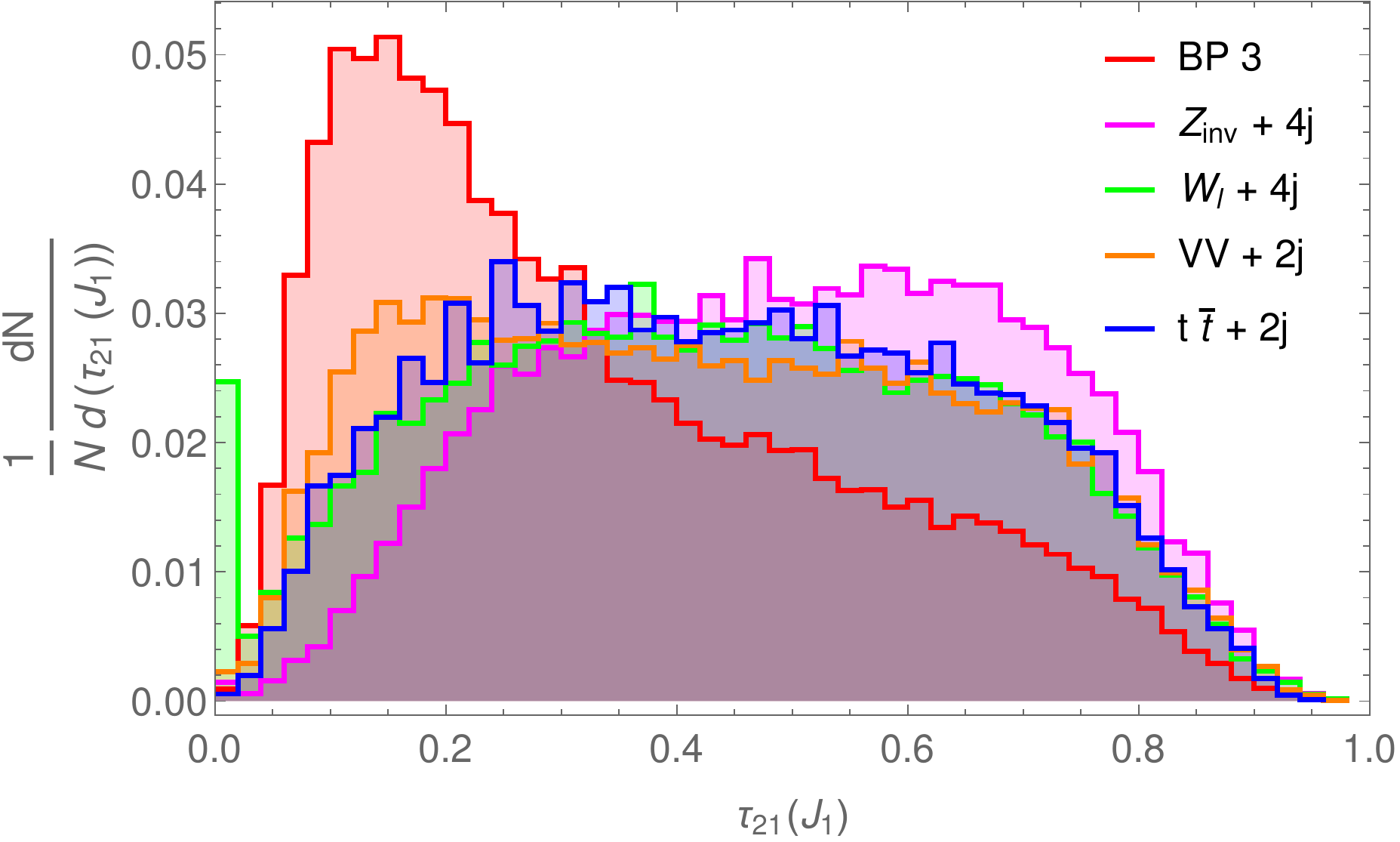}
 	\caption{Normalized distributions for $N$-subjettiness of the leading fatjet $\tau_{21}(J_0)$ (left) and and subleading fatjet $\tau_{21}(J_1)$ (right) after the $baseline~selection~cuts$.}
 	\label{fig:dist_var2}
 \end{figure*}

\vspace{0.2cm}
\noindent
\underline{$N$-subjettiness ratio:}\\
$N$-subjettiness is a jet variable which determines the inclusive jet shape by assuming $N$ subjets in it. 
It is defined as the angular separation of constituents of a jet with the nearest subjet axis weighted by the $P_T$ of the constituents and can be calculated as \cite{Thaler:2010tr,Thaler:2011gf} 
\begin{eqnarray}
\tau_N^{(\beta)} = \frac{1}{\mathcal{N}_0} \sum\limits_i p_{i,T} \min \left\lbrace \Delta R _{i1}^\beta, \Delta R _{i2}^\beta, \cdots, \Delta R _{iN}^\beta \right\rbrace.
\label{eq:nsub_N}
\end{eqnarray}
Here, $i$ runs over the constituent particles inside the jet and $p_{i,T}$ is the respective transverse momentum. The normalization factor is defined as  $\mathcal{N}_0=\sum_{i}p_{i,T} R$ for a jet of radius $R$. In Fig.~\ref{fig:dist_var2}, the distribution for the $N$-subjettiness ratio for signal BP3 and leading background are shown. The value for $\tau_{21}$ is small for fatjets emerging from the signal than the background. The  $N$-subjettiness ratio $\tau_{21}$ is close to zero if correctly identify the $N$-prong structure of the jet.

\subsection{Event selection}
\label{ssec:selec}

We list our baseline selection criteria to select events for further analysis.

\vspace{0.2cm}
\noindent
\underline{Baseline selection criteria:}

\begin{itemize}
\item Events are selected with missing transverse energy $\slashed{E}_T > 100$~GeV.
\item We demand for at least two fatjets of radius parameter $R = 0.8$ constructed using the CA algorithm with
fatjet transverse momentum $P_T(J) > 180$~GeV.
\item We apply the following lepton veto, so that, events are rejected if they contain a lepton with 
transverse momentum $P_T(\ell)  > 10$~GeV and pseudorapidity $|\eta(\ell)| < 2.4$. 
\item We further demand that the azimuthal separation $\Delta\phi$ between the fatjets and $\slashed{E}_T$, 
$|\Delta\phi(J,\slashed{E}_T)| > 0.2$. This minimizes the effect of jet mismeasurement contributing to $\slashed{E}_T$.

\end{itemize}

\noindent
After primary selection, we apply the following final selection criteria on events satisfying the baseline selection criteria for final analysis.

\vspace{0.2cm}
\noindent
\underline{Final selection criteria:}
\begin{itemize}
	
\item After optimization with signal and background, the minimum $\slashed{E}_T$ requirement is raised from 
100 to 200 GeV.

\item  In order to reduce the huge background coming from the $tt+jets$, we apply a $b$-veto with $p_T$-dependent 
$b$-tagging efficiency as implemented in {\tt Delphes}. Here, jets are formed using the anti-$k_t$ algorithm with
radius parameter $R=0.5$.

\item  We demand that the pruned jet mass of leading and subleading fatjets  
should be in $65~\textrm{GeV}<M_{J_i}<105~\textrm{GeV}$ to tag $J_V$ candidates. 
 
\item Further to discriminate the fatjet $J_V$ from the QCD jets, we look for the two-prong nature of the fatjet using 
$N$-subjettiness and select the events with $\tau_{21}(J_i) < 0.35$ of the unpruned fatjet.  

\end{itemize}
 
 %%%%%%%%%%%%%%%%%%%%%%%%% scale=0.715
 %%%%%%%%%%%%%%%%%%%%%%%%%
 
\begin{table*}[htb!]
\tiny	
\centering
\renewcommand{\arraystretch}{2.5}
\begin{tabular}{|c||c|c||c|c|c|}
	\hline
	Cut    &  \multicolumn{5}{c|}{Signal  \bfseries{BP3}}                                 \\ \hline\hline
	& $ AH^{0}$ & $ H^{\pm}H^{0}$              & $ AH^{\pm}$ & $ H^{+}H^{-}$ & $AA $
	\\ \hline
	Baseline $+\slashed{E}_T > 200$~GeV      
	& \multirow{2}{*}{672.03 } & \multirow{2}{*}{1608.8} & \multirow{2}{*}{711.62} &  \multirow{2}{*}{562.15}& \multirow{2}{*}{64.5}
		
		\\   & [$100\%$] &[$100\%$]&[$100\%$]&[$100\%$]& [$100\%$]

		\\ \hline $b$-veto   & 474.24  & 1291.74  & 474.81  & 426.85 & 32.5

		\\   &[$70.70\%$]        &[$80.28\%$]         &[$66.66\%$]       &[$75.95\%$]  &   [$50.49\%$] 
		
		\\\hline
	$65~\textrm{GeV} < M(J_0),M(J_1) < 105~\textrm{GeV}$ & 79.50  & 274.17 & 171.83 & 137.56 & 4.87

		\\   & [$11.83\%$] &[$17.04\%$] &[$24.12\%$]&[$24.48\%$]  &   [$18.13\%$]

		\\\hline
		$\tau_{21}(J_0),\tau_{21}(J_1) < 0.35$ & 52.44  & 171.79  & 128.40  & 101.98 & 3.5 
		
		\\ &[$7.88\%$]&[$10.67\%$]&[$18.02\%$]&[$18.13\%$]    &  [$5.18\%$]        
			
		\\\hline
	\end{tabular}

\caption{After implementing the corresponding cut, the expected number of events and cut efficiency are shown for  signal ({\tt BP3}) for all possible channels which are contributing to the two $2J_V$ +  MET final state,for an integrated luminosity of $3000$ fb$^{-1}$ at the 14 TeV LHC.} 
	
	\label{tab:cut_flow_sig}
\end{table*}

%======================================================

\begin{table*}[htb!]
	\tiny	
	\centering
	\renewcommand{\arraystretch}{2.5}
	\begin{tabular}{|c||c|c|c|c|c|c|}
		\hline
		Cut    &  \multicolumn{5}{c|}{Background}                                 \\ \hline\hline
	 & $Z_{\nu} + jets $ & $W_\ell + jets $ & $VV+jets$ & $Single-top$ & $tt+jets$
		\\ \hline
		Baseline $+\slashed{E}_T > 200$~GeV      
		& \multirow{2}{*}{$3.22\times 10^6$} &  \multirow{2}{*}{$4.76\times 10^6$} & \multirow{2}{*}{$1.47\times 10^5$ } & \multirow{2}{*}{$2.06\times 10^5$} & \multirow{2}{*}{$3.81\times 10^5$} 
		
		\\   & [$100\%$] &[$100\%$]&[$100\%$]&[$100\%$]&[$100\%$]

		\\ \hline $b$-veto & $2.69\times 10^6$   & $4.30\times 10^5$   & $1.13\times 10^5$     &  $3.63\times 10^4$   &  $4.60\times 10^4$

		\\   &[$83.65\%$]       &[$9.22\%$]        &[$75.01\%$]        &[$16.90\%$]   &[$12.34\%$]       
		
		\\\hline
		$65~\textrm{GeV} < M(J_0),M(J_1) < 105~\textrm{GeV}$	& $3.80\times 10^4$ &  $1.67\times 10^4$  & $4.09\times 10^3$ & $1.96\times 10^3$ & $1.66\times 10^3$

		\\  &[$1.19\%$]&[$0.35\%$]    & [$2.81\%$] &[$0.72\%$]   &[$0.43\%$]

		\\\hline
$\tau_{21}(J_0),\tau_{21}(J_1) < 0.35$ 	  & $1.30\times 10^4$ &$3.79\times 10^3$ & $1.62\times 10^3$  & $1.44\times 10^3$  & $3.84\times 10^2$ 
		
		\\&[$0.41\%$]&[$0.07\%$]   &[$1.03\%$]&[$0.56\%$] &[$0.10\%$]              
		
    	\\\hline
	\end{tabular}
	\caption{Cut flow for the SM backgrounds after corresponding cuts are implemented, for an integrated luminosity of $3000$ fb$^{-1}$ at the 14 TeV LHC.}
	\label{tab:cut-flow-bg}
\end{table*}

%====================================================== = = = = = = = = = = 

In Table~\ref{tab:cut_flow_sig}, we present the cut-flow for the signal (BP3) associated with the cut efficiencies and the 
number of events for an integrated luminosity of 3000~fb$^{-1}$ at the 14 TeV LHC. Similarly, Table~\ref{tab:cut-flow-bg} represents  
the cut-flow for the different backgrounds. From these numbers, it is explicit 
that the $\tau_{21}$ and $M_J$ are powerful variables to have large background reduction with good signal acceptance.
We can further infer from Table~\ref{tab:cut_flow_sig} that in spite of quite low efficiencies of $AH$ and $H^{\pm}H$ channels to satisfy the $2 J_V+\slashed{E}_T$ criteria, they give comparable contributions 
to the signal due to its large production cross section. 
 
We compute the statistical signal significance using $\mc{S}=\mathcal{N}_{S}/\sqrt{\mathcal{N}_{S}+\mathcal{N}_{B}}$, where $\mathcal{N}_{S}$ and $\mathcal{N}_{B}$ represent the remaining number of signal and background events after implementing  all  the cuts. We show the statistical significance for different benchmark points in Table~\ref{tab:Significance}. The highest significance is found for BP3. We would like to emphasize that even after utilizing the novel techniques of jet substructure this particular region of parameter space is very challenging to probe with high sensitivity at the HL-LHC. In order to optimize our search further, we use MVA with jet substructure variables.

\begin{table}[h]
	\centering
	\renewcommand{\arraystretch}{1.5}
	\begin{tabular}{|c|c|c|c|c|c|c|c|}
		\hline
		Benchmarks   & BP1  & BP2 & BP3 & BP4 & BP5  & BP6 & BP7\\
		\hline
		\hline
		Significance &  1.9	& 2.9  & 3.2& 2.9  & 1.9 & 1.6& 1.1 \\ 
    	\hline
	\end{tabular}
	\caption{ Statistical  significance  of  the  signal  for  different  benchmark  points  in  di-fatjet + $\slashed{E_T}$ analysis for an integrated luminosity of $3000$ fb$^{-1}$ at the 14 TeV LHC. 
	}
	\label{tab:Significance}
\end{table}

%%%%%%%%%%%%%%%%%%%%%%%%%%%%%%%%%%%%%%%%%%%%%
%%%%%%%%%%%%%%%%%%%%%%%%%%%%%%%%%%%%%%%%%%%%%%%%
\section{Multivariate analysis}
\label{sec:mva}

In the previous section, we present the reach of our model using a CBA. Although we have not achieved discovery significance of $5\sg$ in any of our benchmark points, we see that the two variables viz. $M_{J}$ and $\tau_{21}$ are very powerful to separate the tiny signal from the large SM background. In this section, we use a sophisticated MVA to achieve better sensitivity than a CBA. We would like to discuss two important points here. First, we have observed that MVA does not perform well if we use events selected just with the baseline cuts since the signal is too tiny compared to the overwhelmingly large background. Therefore, we need to apply, in addition to the baseline selection cuts, the following strong cut on the hardest fatjet mass, $M_{J_0}>40$ GeV and b-veto on jets to further trim down the large background before passing events to MVA. These cuts are very effective to drastically reduce the background but not the signal and are optimally
chosen such that it is not too close or too relaxed compared to the cuts used in CBA. If the extra strong cuts for MVA are too close to the cuts applied for the CBA, MVA will not give us an improved sensitivity. On the other hand, if they are too relaxed, the performance of MVA will degrade as the background will become too large. Although we select events with two high-$p_T$ fatjets, we only demand the jet mass
of the leading-$p_T$ fatjet is greater than 40 GeV. This will pass a large fraction of mono-fatjet signal events along with the di-fatjet. Therefore, on the one hand, this will increase the signal. However, on the other hand, this will also increase the background. 

\begin{table}[!htbp]
\small
\centering
\begin{tabular}{cccccccc}

\hline \hline
Topology & BP1 & BP2 & BP3 & BP4 & BP5 & BP6 & BP7 \\ 
\hline 
$1 J_V$ & 1668 & 2025 & 2023 & 1472 & 1334 & 1190 & 920 \\ 
$2 J_V$ & 601 & 1112 & 1572 & 1254 & 979 & 948 & 608 \\ 
\hline
\hline
Z & W & t & tt & WZ & ZZ & WW & Total \\ 
\hline 
$3.15\times10^6 $ & $1.43\times10^6 $ & $1.6\times10^5 $ & $1.6\times10^5 $ & $1.76\times10^5 $ & $2.97\times10^4 $ & $1.21\times10^4 $&$5.1\times10^6 $ \\
\hline 
\end{tabular} 
\caption{Number of signal and background events at the 14 TeV LHC with 3000 fb$^{-1}$ integrated luminosity. These numbers are
obtained by applying $M_{J_0}>40$ GeV and $b$ - jet veto in addition to the baseline cuts defined in the text.}
\label{tab:event}
\end{table}

In Table~\ref{tab:event}, we show the number of signal ($1 J_V$ and $2 J_V$ categories) and background events at the 14 TeV
LHC with 3000 fb$^{-1}$ integrated luminosity. Observe that although we demand two fatjets in our selection, the number of $1 J_V$ events that contribute to the signal are always bigger than the $2 J_V$ contributions for all BPs. This is due to the fact that
cross sections for $1 J_V$ topologies are much bigger than the $2 J_V$ topologies and also a significant fraction of $1 J_V$ events
pass the selection cuts. Therefore, it is necessary to design hybrid selection cuts, stricter than $1 J_V$ but
looser than $2 J_V$, where both $1 J_V$ and $2 J_V$ topologies contribute. Our selection cuts are, therefore, optimally designed 
to achieve better sensitivity.

For our MVA, we use the adaptive BDT algorithm. 
We obtain two statistically independent event samples for the signal as well as for the background and split the dataset randomly $50\%$ for testing and the rest for training purposes for both the signal and background. Note that there are multiple processes that are contributing to the signal and similarly for the background. In MVA, we construct the signal classes by combining both the $1 J_V$ and $2 J_V$ topologies that pass our MVA selection criteria. These different signal samples are separately generated at LO and then mixed 
according to their proper weights to obtain the kinematical distributions for the combined signal. Similarly, all different
background samples are mixed to obtain similar distributions for the background class.

The final set of variables that are used in the MVA are decided from
a larger set of kinematic variables by looking at their power of discrimination between signal and background classes. Four substructure variables for two fatjets, {\emph i.e.}, $M_{J_{0,1}}$ and  $\tau_{21}(J_{0,1})$ has already proved to be very important discriminators in our CBA. Stronger transverse momenta cut for such jets are favorable to retain the correct classification of these variables. We already required  reasonably high $P_T$ criteria for both such jets. However, to construct the hybrid selection cuts $P_T(J_0)$ can still take a role in determining the purity of the hardest fatjet $J_0$. We also include relative separation between these fatjets $\Delta R (J_0, J_1)$ and the azimuthal angle separation between the leading fatjet from the missing transverse momentum direction $\Delta \phi (J_0, \slashed{E}_T)$.  
The scale of new physics is relatively high, and that is typically captured by some of the topology independent inclusive variables like $H_T$, $\slashed{H}_T$, $\slashed{E}_T$ etc. We utilize the global inclusive variable $\sqrt{\hat{S}_{min}}$ proposed to determine the mass scale of new physics for events with invisible particles such as ours~\cite{Konar:2008ei,Konar:2010ma,Barr:2011xt}. This variable, constructed out of all reconstructed objects at the detector, demonstrates better efficiency compared to other inclusive counterparts. For example, we did not use $\slashed{E}_T$ as a feature after baseline cut since it showed a high correlation with $\sqrt{\hat{S}_{min}}$ and  turned out to be less important than it.

%===========================================================
\begin{figure}[t!]
	\centering
	\includegraphics[scale=0.28]{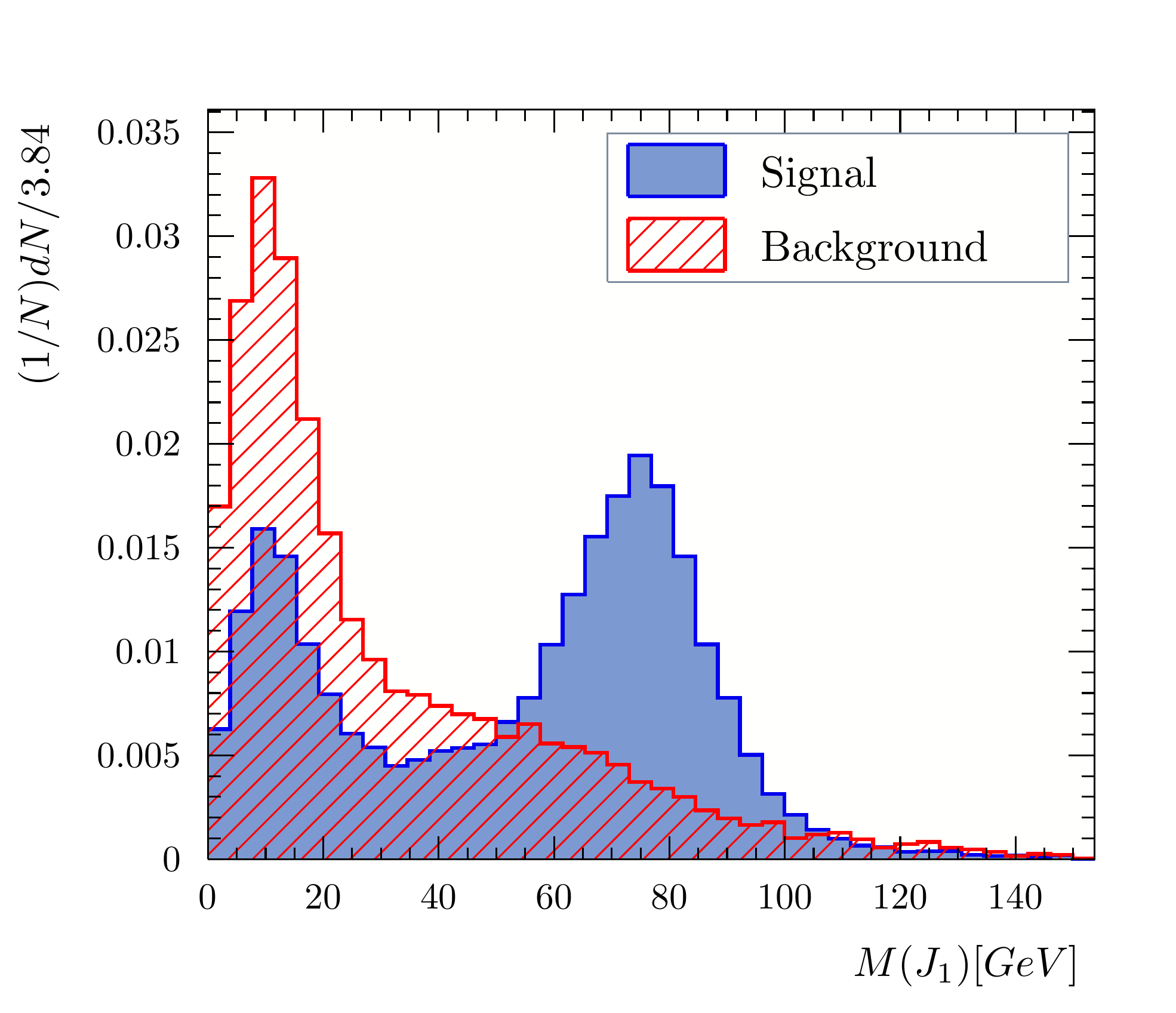}~~~~
	\includegraphics[scale=0.28]{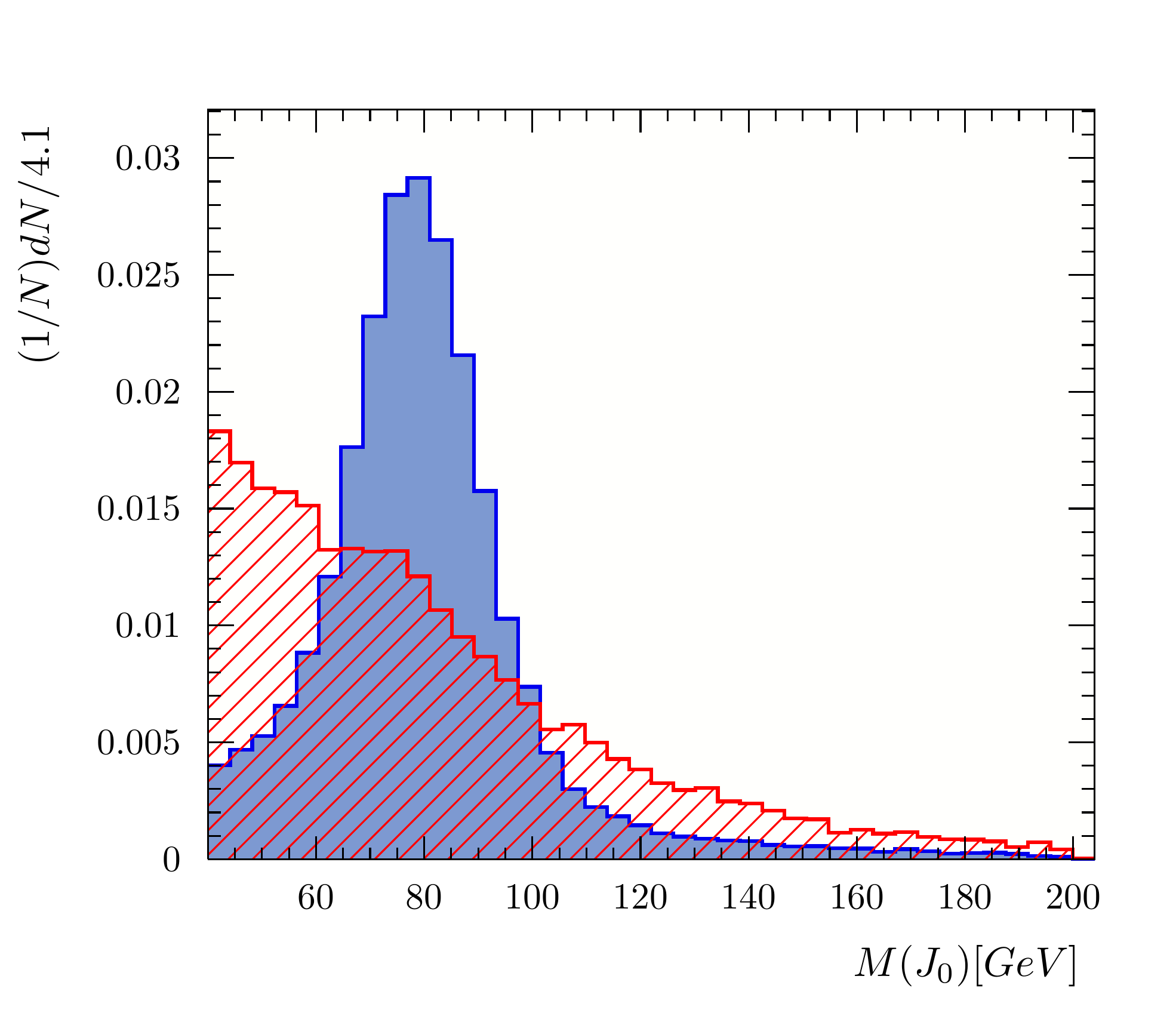}\\
	\includegraphics[scale=0.28]{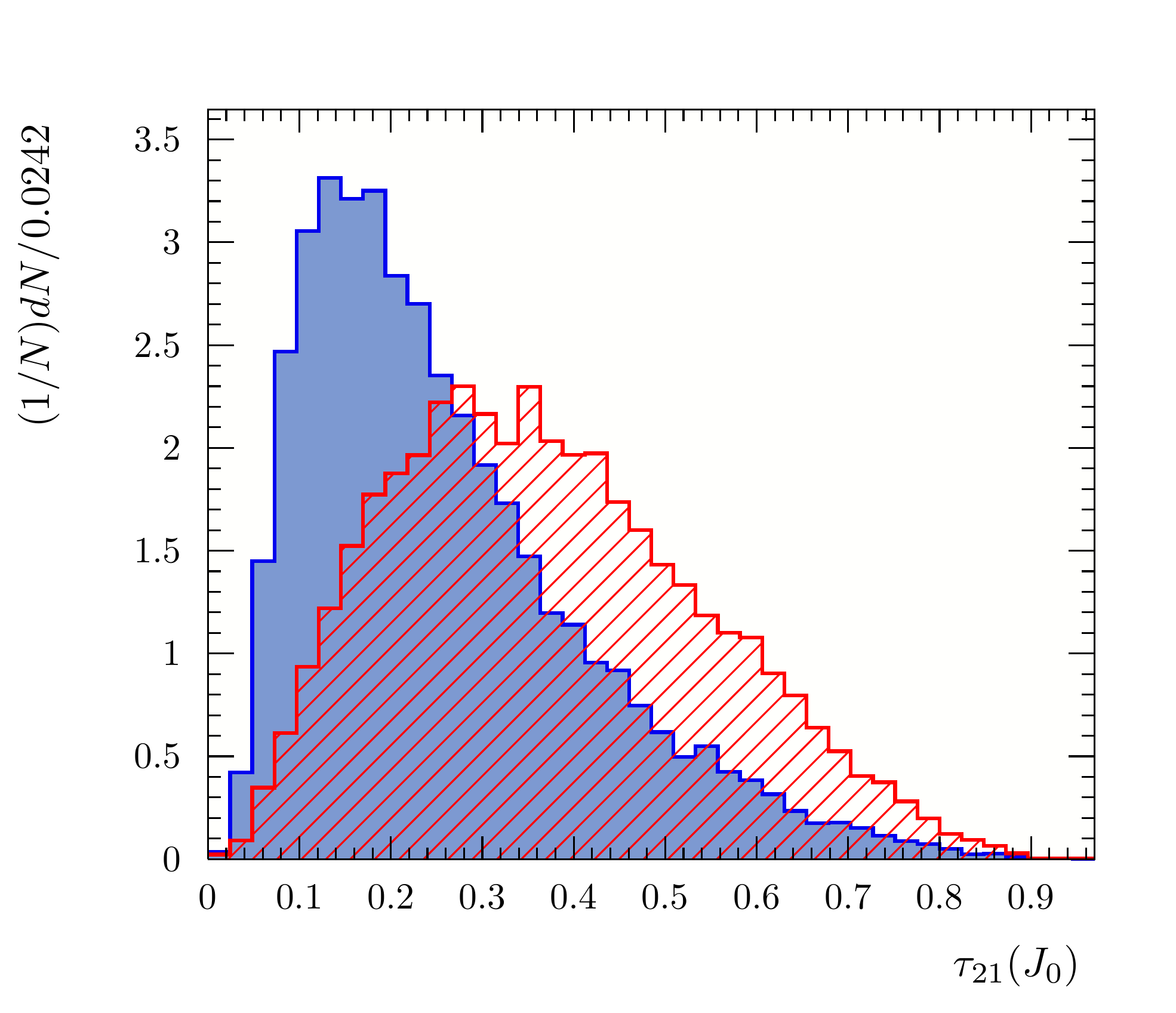}~~~~
	\includegraphics[scale=0.28]{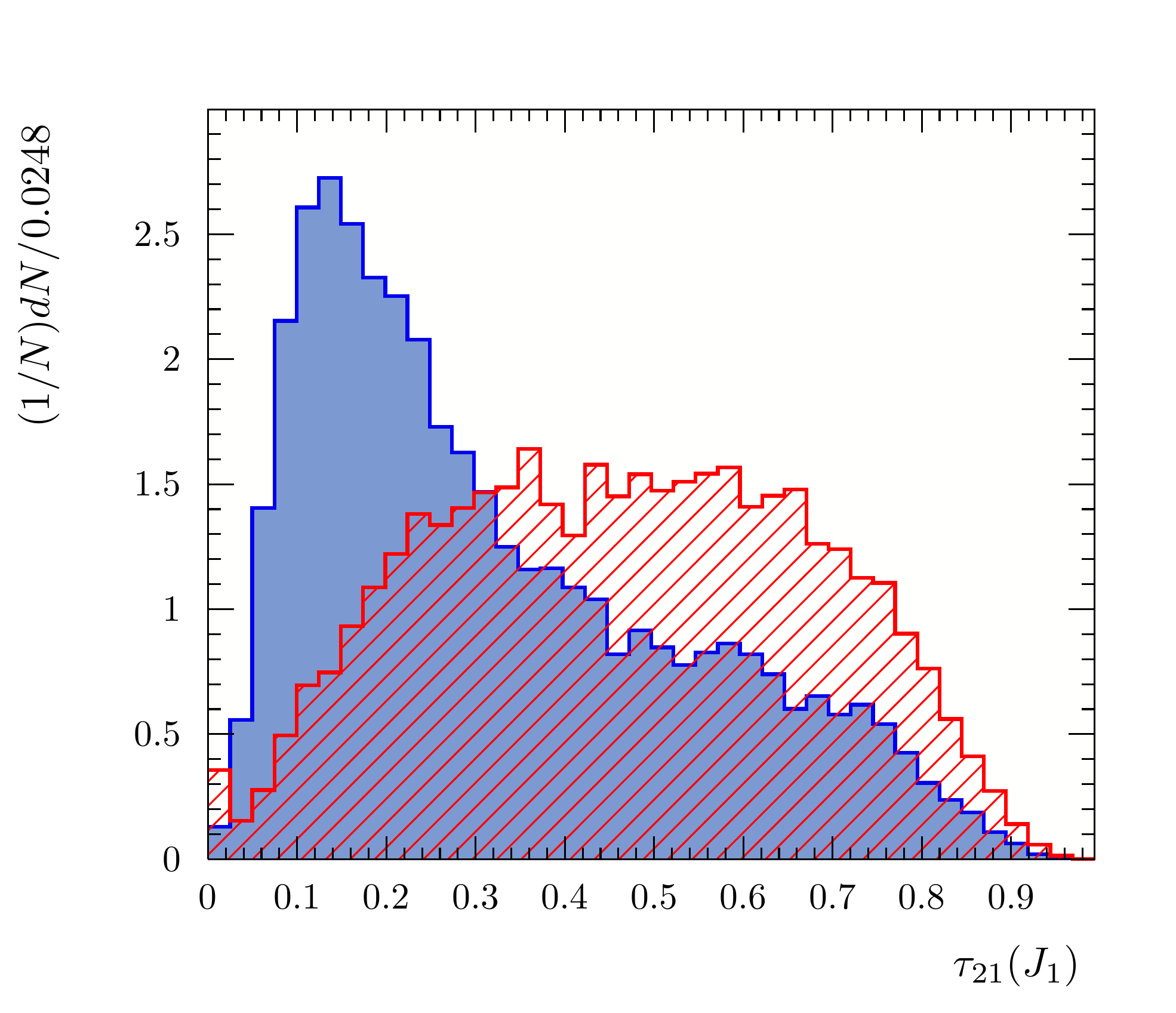}\\
	\includegraphics[scale=0.28]{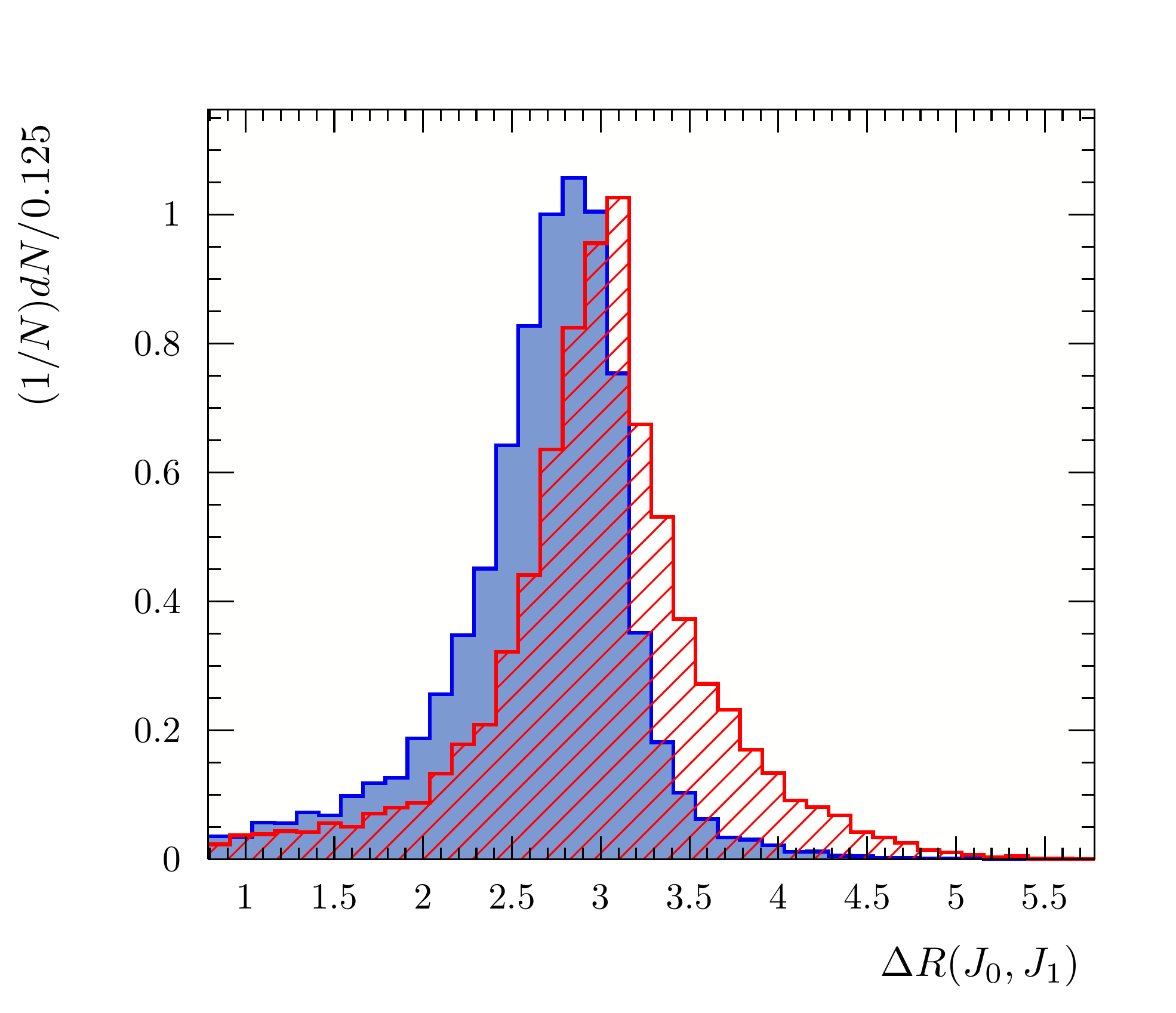}~~~~
	\includegraphics[scale=0.28]{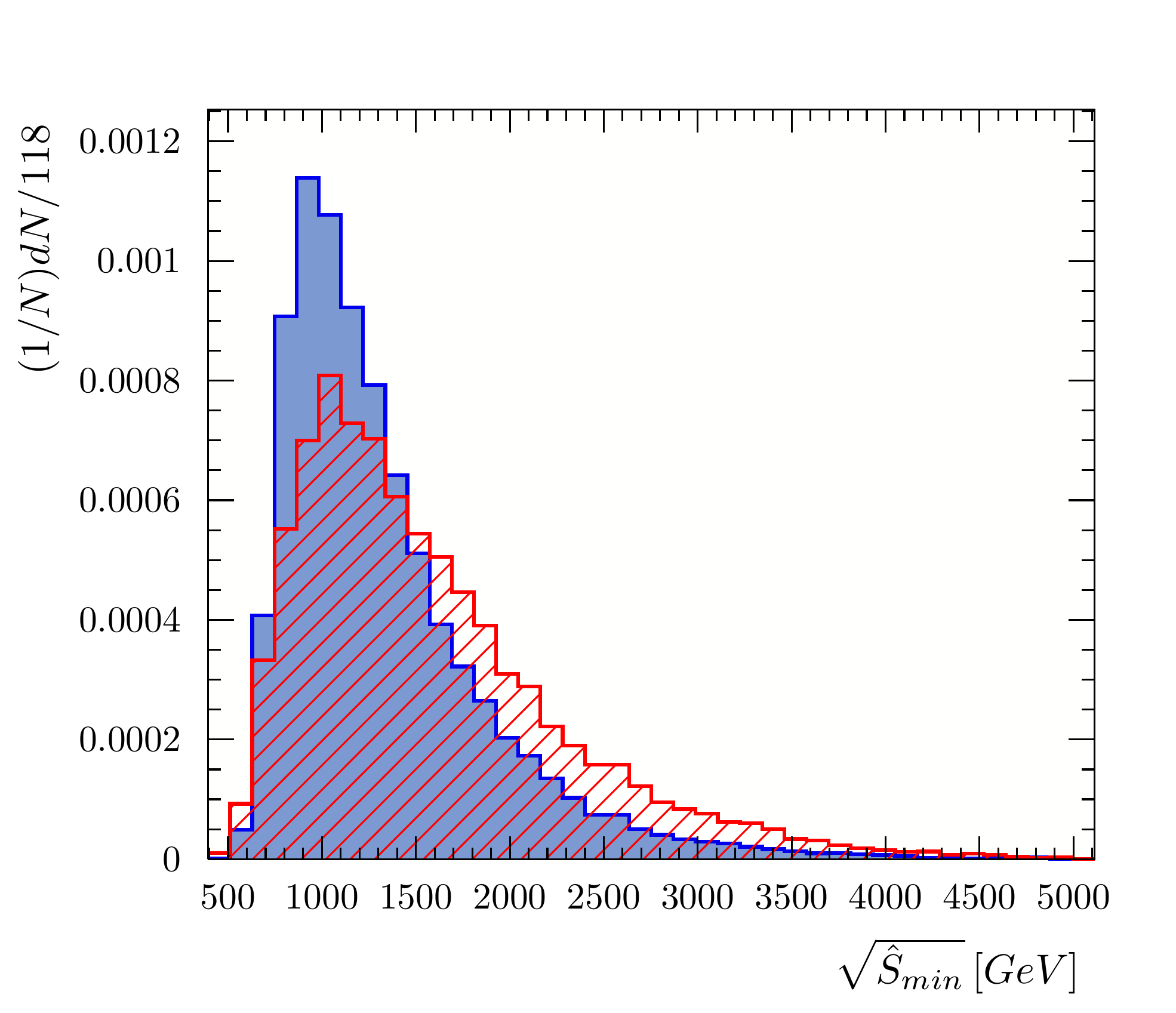}\\
	\includegraphics[scale=0.28]{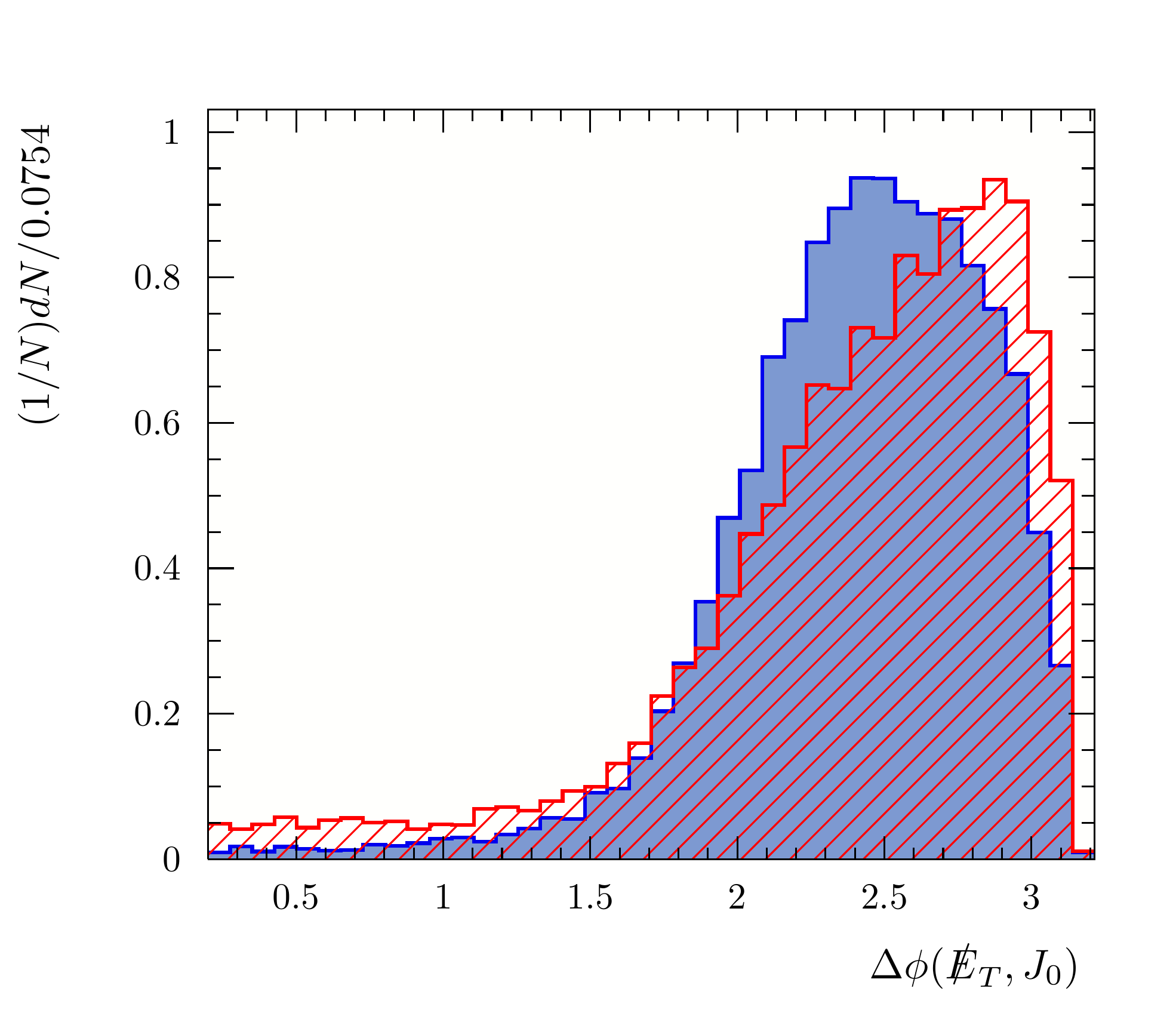}~~~~
	\includegraphics[scale=0.28]{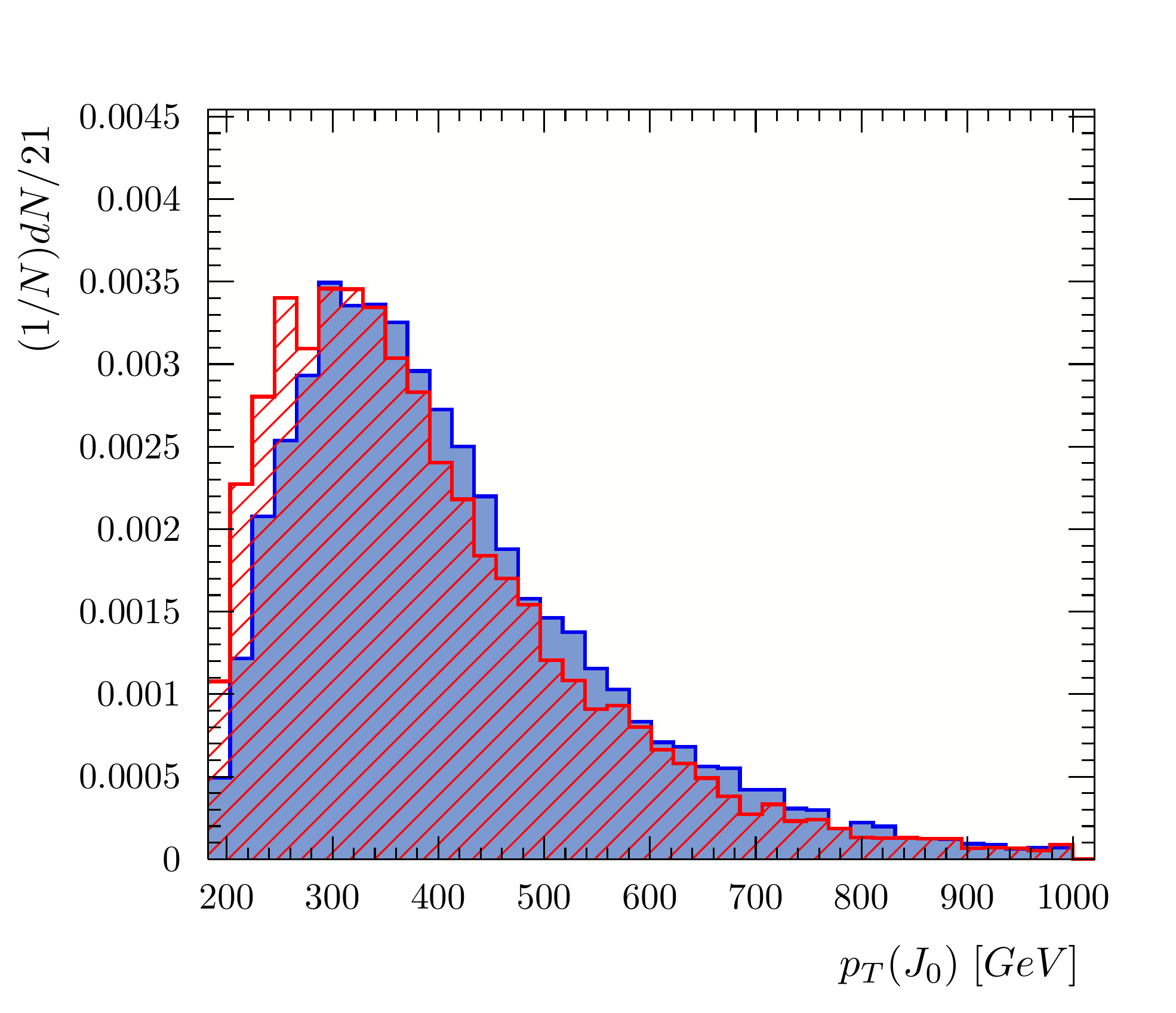}
	\caption{Normalized distributions of the input variables at the LHC ($\sqrt{s}=14$ TeV) used in the MVA for the signal (blue) and the background (red). Signal distributions are obtained for BP3 including $1 J_V$ and $2 J_V$ topologies and the background includes all the dominant backgrounds discussed in Sec.~\ref{subsec:bg}.}
	\label{fig:varBP3}
\end{figure}
%===========================================================

%===========================================================
\begin{figure}[t!]  
\centering
\includegraphics[height=5cm,width=8.5cm]{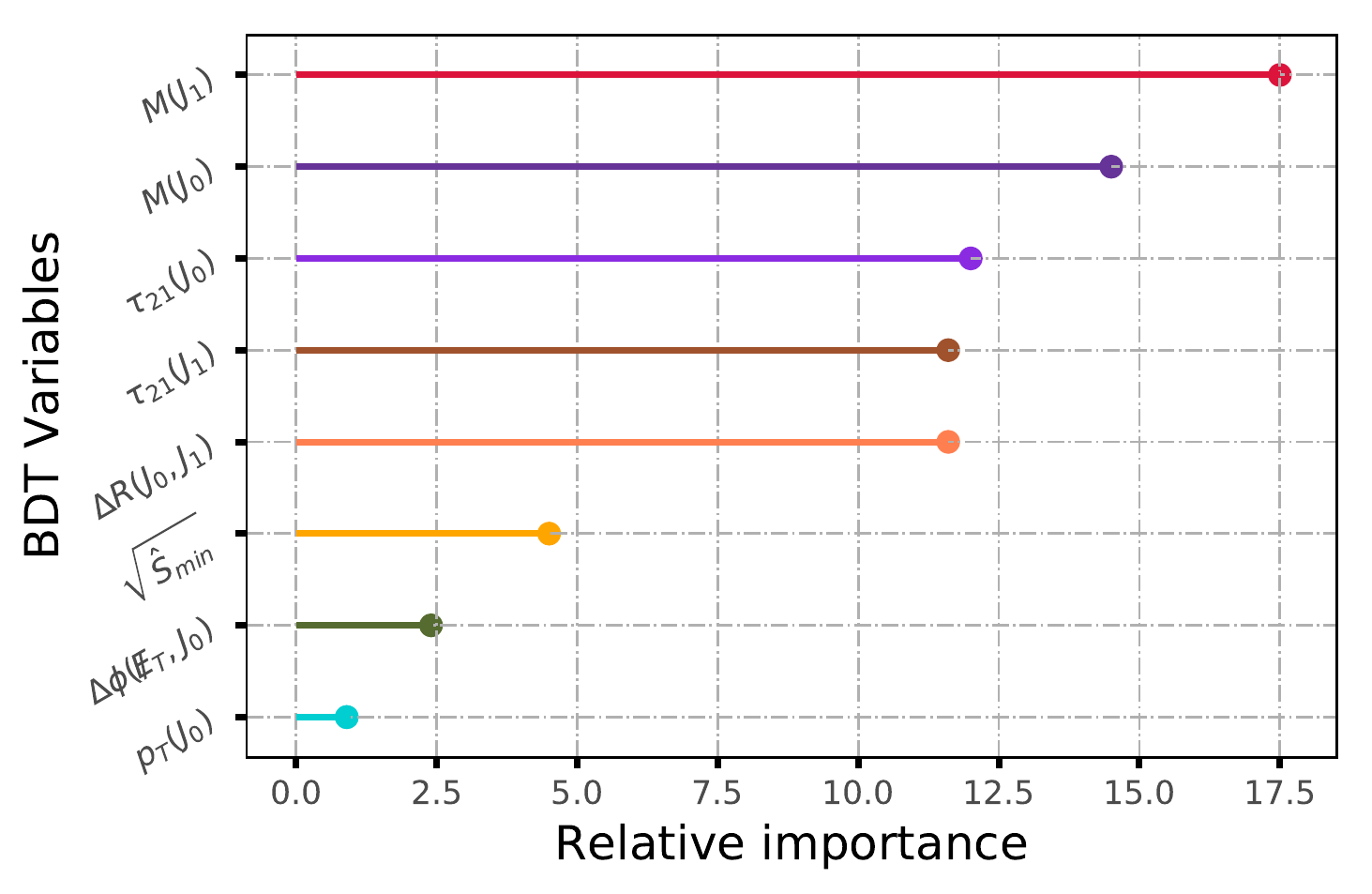}
\caption{Kinematic variables used for our MVA and their relative importance. We obtain these numbers from the TMVA package 
for the benchmark point BP3. Here, we show method unspecific relative importance. This can change slightly for different
algorithms and their tuning parameters.}
\label{fig:Var_Imp}
\end{figure}
%===========================================================

\begin{figure}[t!]  
\centering
\includegraphics[width=0.5\textwidth]{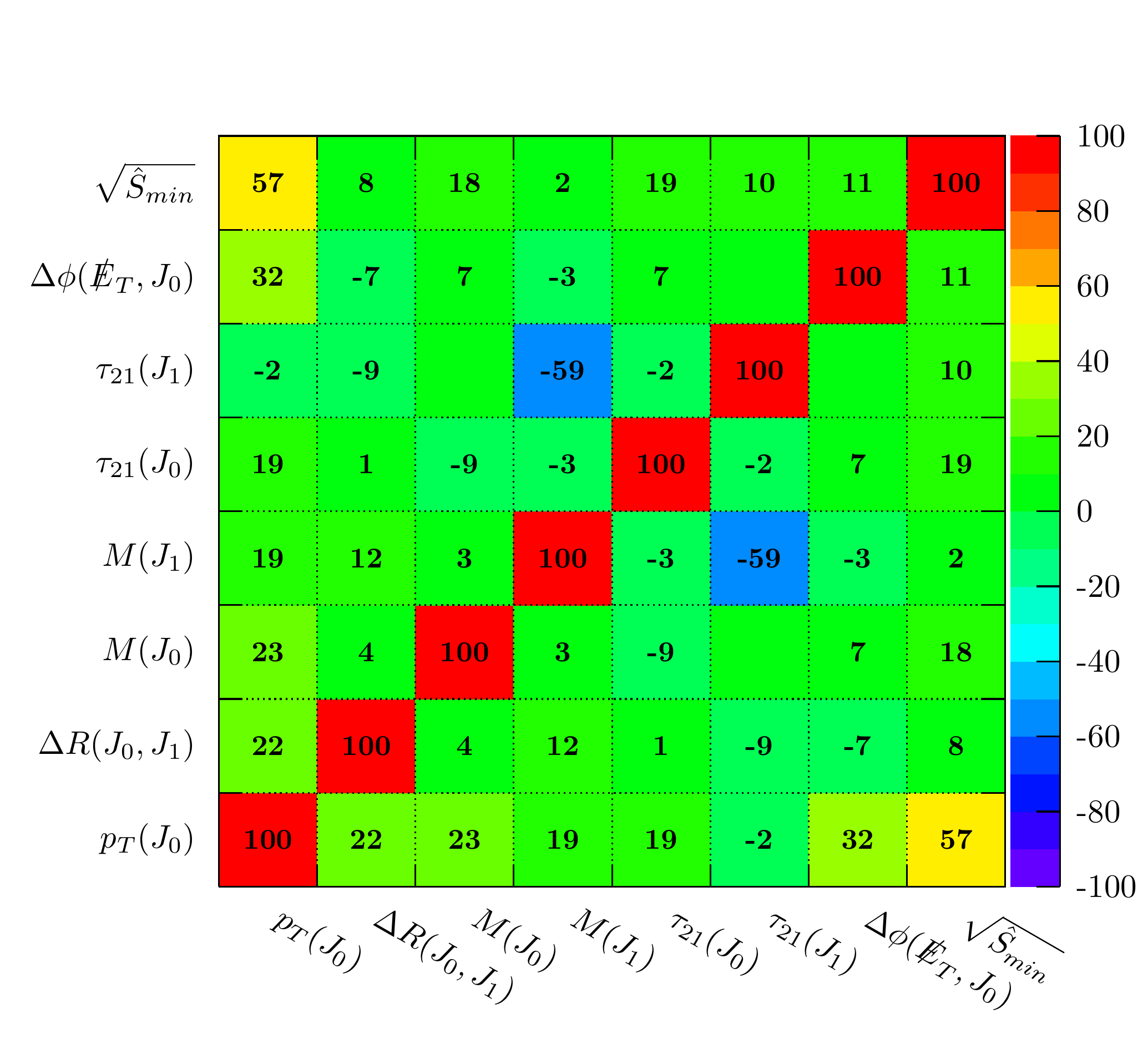}~~
\includegraphics[width=0.5\textwidth]{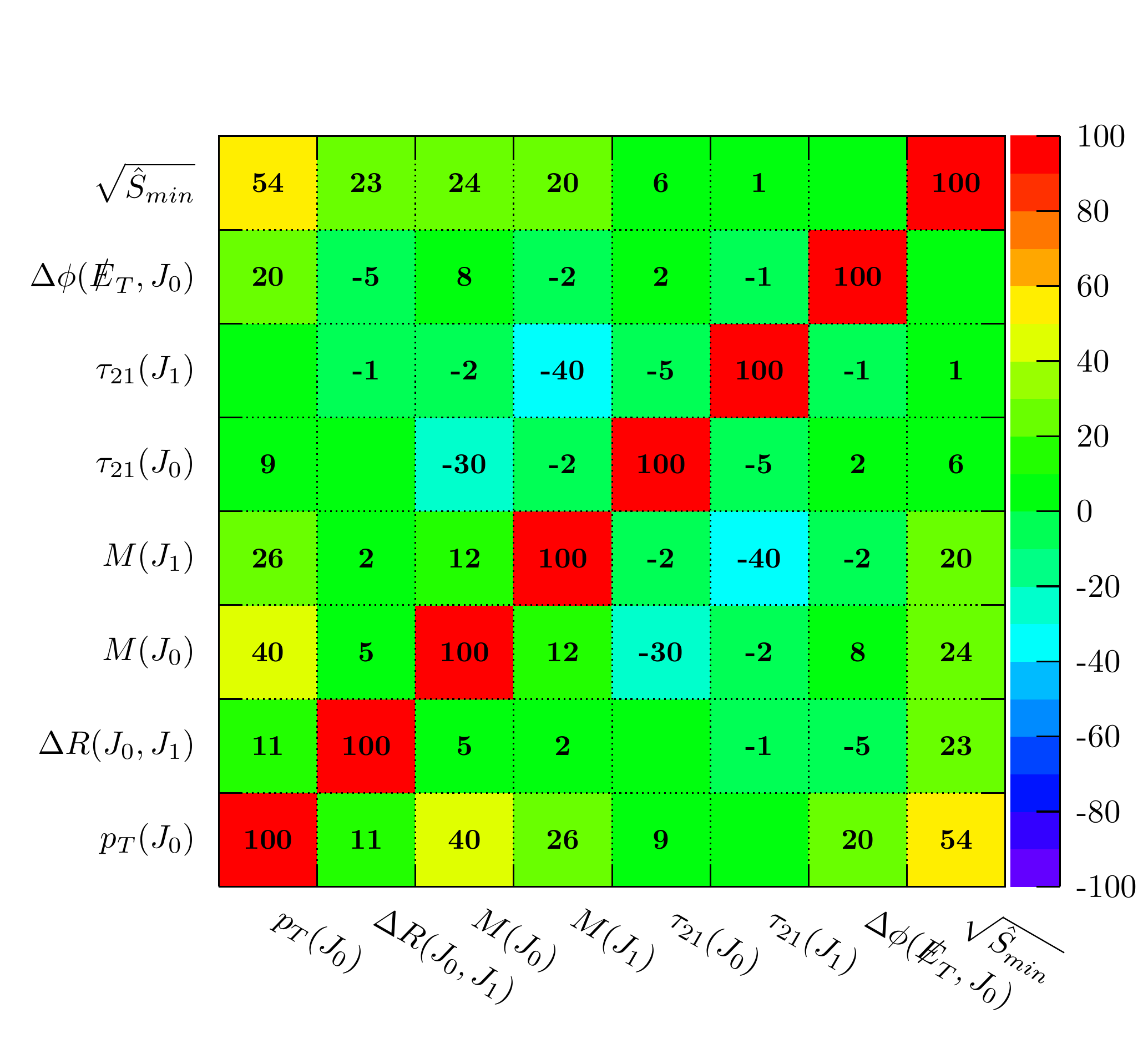}
\caption{The linear correlation coefficients (in \%) for signal (left panel) and background (right panel) among different kinematical variables that are used for the MVA for BP3. Positive and negative signs of the coefficients signify that the two
variables are positively correlated and negatively correlated (anti-correlated).}
\label{fig:corr}
\end{figure}

In Fig.~\ref{fig:varBP3}, we show the normalized distributions of all eight input variables used in the MVA. Signal distributions are obtained for BP3 including $1 J_V$ and $2 J_V$ topologies and the background includes all the dominant backgrounds discussed in Sec.~\ref{subsec:bg} for the 14 TeV LHC. For the same benchmark scenario, method unspecific relative importance of all the kinematic variables are available during TMVA analysis and presented in   Fig.~\ref{fig:Var_Imp}.
 Moreover, we mostly keep variables which are less correlated (or anti-correlated) for both the signal and the background.
Relative importance is a measure that is used to rank the variables in MVA. In other words, a variable has better discriminatory power if it has greater relative importance. For this particular benchmark point, BP3, $M_{J_{0,1}}$ 
variables are very good discriminators according to their relative importance. The $N$-subjettiness variables, $\tau_{21}(J_{0,1})$, are
also very good discriminators as expected. Note that, the relative importance can change for different benchmark points or different LHC energies etc., that can change the shapes of the variables. The linear correlation matrices for the signal and the background can be seen in Fig.~\ref{fig:corr}.
Observe that $M_{J_1}$ and $\tau_{21}(J_1)$ variables are strongly anti-correlated. The correlation in the $M_{J_1}$ and $\tau_{21}(J_1)$ variable is due to a mixture of $1J_V$ and $2J_V$ topology in the signal.  However, we keep both of them in the MVA since both of them are very powerful discriminators for $2J_V$ topology.

%===========================================================
%===========================================================

Since the BDT algorithm is prone to overtraining, one should be careful while using it. This usually happens during the
training of the algorithm due to inappropriate choices of the BDT specific parameters. 
One can avoid overtraining by checking the Kolmogorov-Smirnov probability during training. 
We train the algorithm for every benchmark point separately and ensure that the algorithm is not overtrained in our analysis.
Note that the set of eight variables that are used in our analysis may not be the optimal ones. There is always the 
scope of improving the analysis by choosing a cleverer set of variables. But since the variables we use in MVA are very
good discriminators, our obtained sensitivities are fairly robust. 

\begin{figure}[t!]  
\centering
\includegraphics[height=6cm,width=7.5cm]{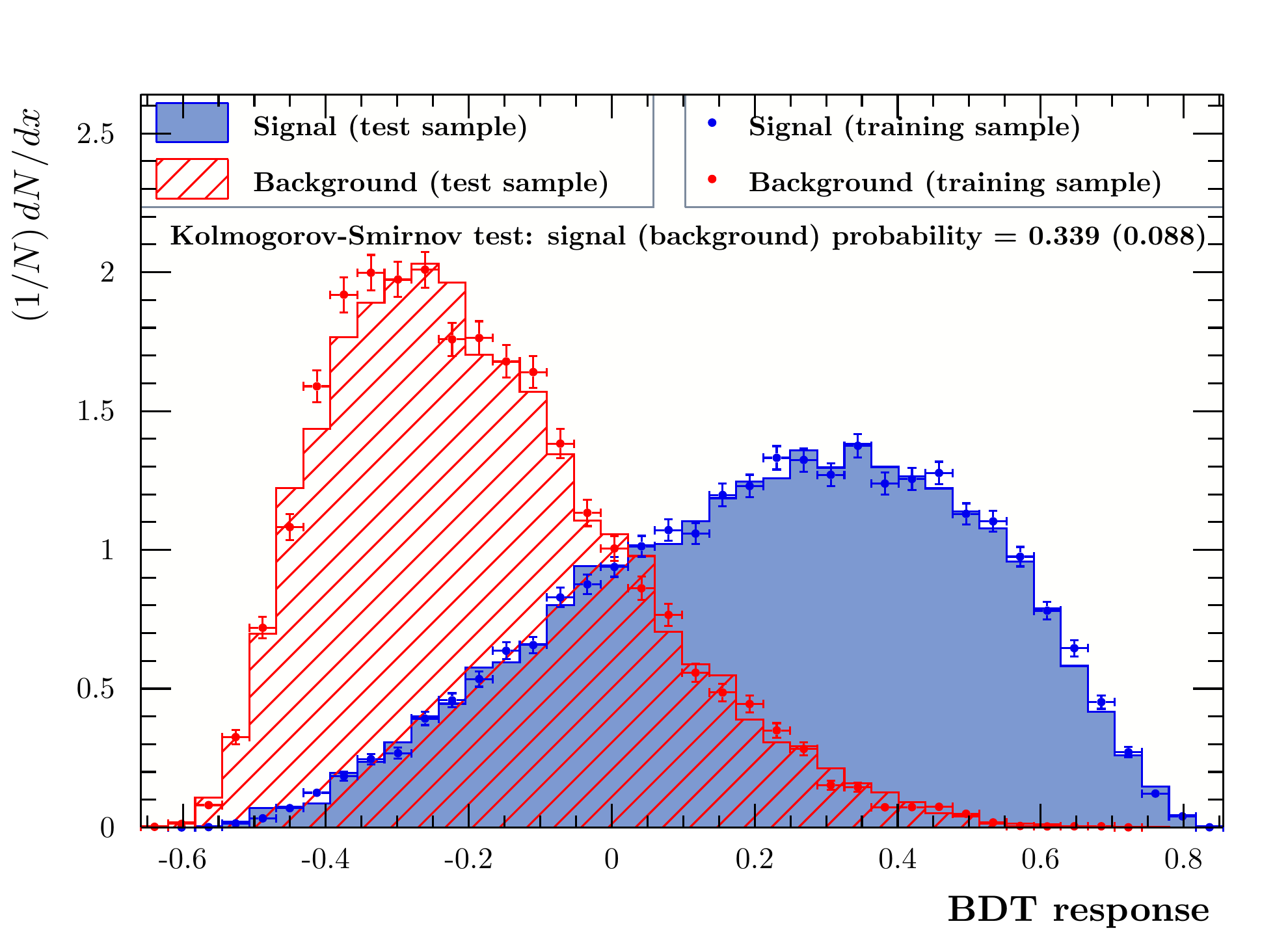}~~
\includegraphics[height=6cm,width=7.5cm]{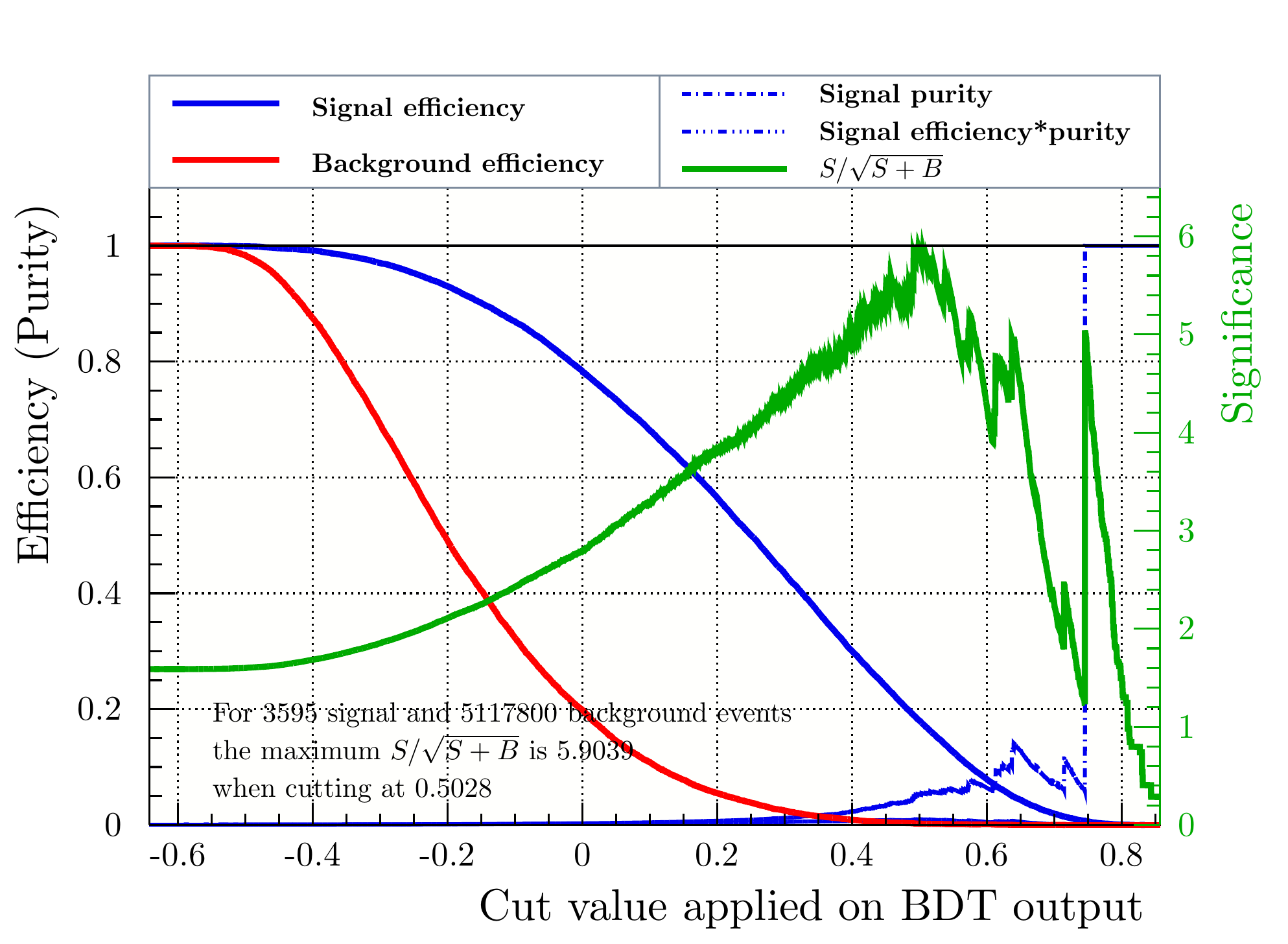}

\caption{(Left panel) Normalized BDT response distributions for the signal and the background for BP3.
(Right panel) Cut efficiencies as functions of BDT cut values.}
\label{fig:BDT}
\end{figure}

\begin{figure*}[t!]
\centering
\includegraphics[height=5cm,width=8cm]{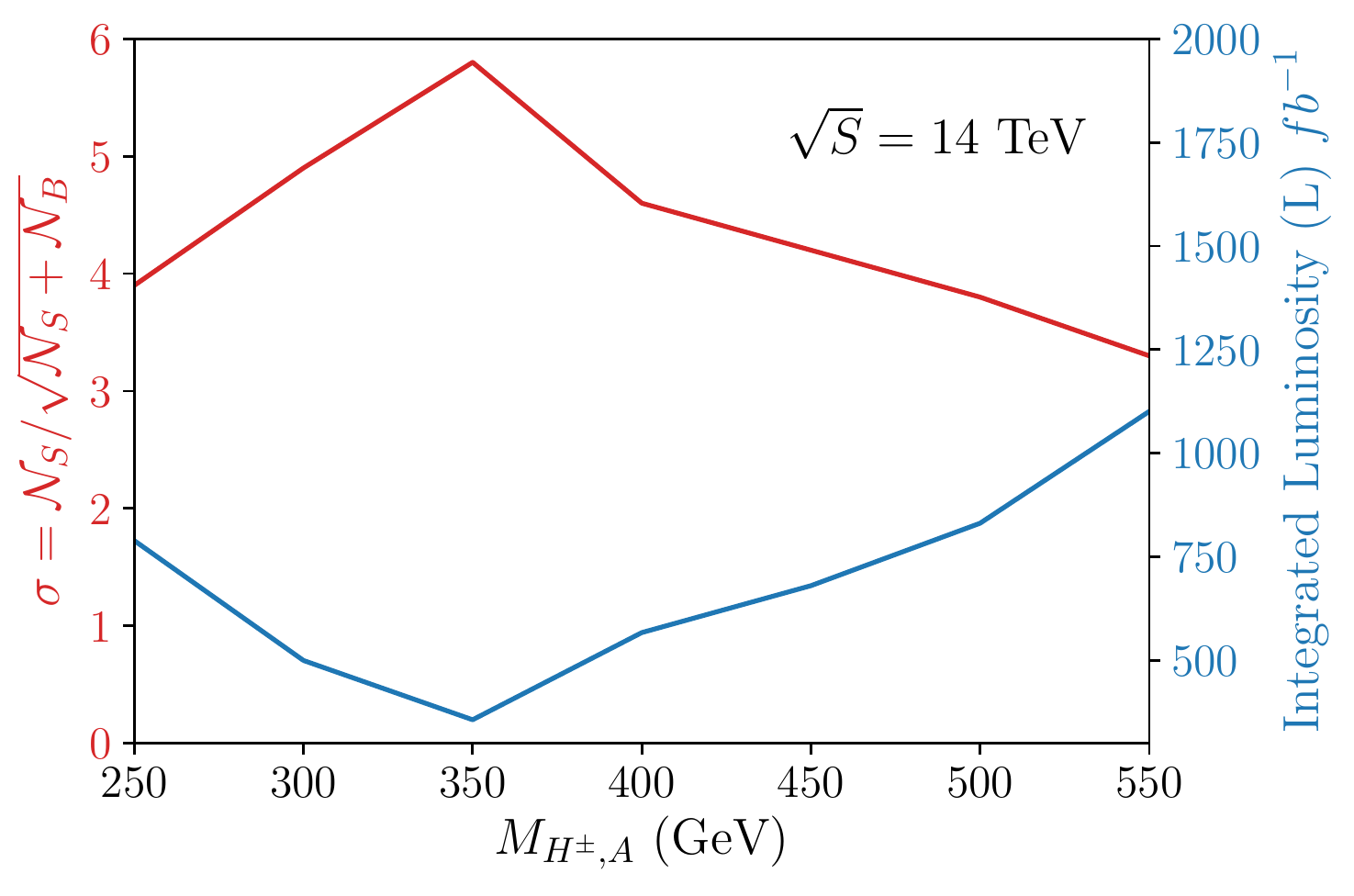}
\caption{Significance as a function of heavy scalar mass $M_{H^{\pm}}$ at the 14 TeV LHC with 3000 fb$^{-1}$ integrated luminosity. We also present required luminosity for the exclusion ($2 \sigma$) of different benchmark points based on this heavy scalar mass.}
\label{fig:signi}
\end{figure*}

In Fig.~\ref{fig:BDT}, we show the normalized BDT response for the signal and the background (training and test samples for both the classes) for BP3. One can clearly see that the BDT responses for the signal and background classes are well separated.
We apply a cut on the BDT responses {\emph i.e.}, $\mathrm{BDT}_{res}>\mathrm{BDT}_{cut}$ and show the corresponding cut efficiencies
for the signal (blue) and the background (red) and the significance (green) as functions of $\mathrm{BDT}_{cut}$.
The significance is computed using the formula $\sg=\mathcal{N}_{S}/\sqrt{\mathcal{N}_{S}+\mathcal{N}_{B}}$ where
$\mathcal{N}_{S}$ and $\mathcal{N}_{B}$ are the signal and background events that are survived after the $\mathrm{BDT}_{res}>\mathrm{BDT}_{cut}$ cut for a given integrated luminosity. The optimal BDT cut, $\mathrm{BDT}_{opt}$ is the cut for which the significance is maximized. In  Table~\ref{tab:BDT}, we show $\mathcal{N}_{S}$, $\mathcal{N}_{B}$ and $\sg$ for different BPs for the 14 TeV LHC, considering  an integrated luminosity of 3000 $fb^{-1}$. We also demonstrate this significance as a function of $M_{H^{\pm},A}$  in Fig.~\ref{fig:signi} (red curve), whereas the blue curve represents the required luminosity for the $2 \sigma$ exclusion of different BPs.

\begin{table}[!htbp]
\centering
\begin{tabular}{c|ccccc}
\hline \hline
BP & $\mathcal{N}_S^{bc}$ & $\mathrm{BDT}_{opt}$ & $\mathcal{N}_{S}$ & $\mathcal{N}_{B}$ & $\mathcal{N}_{S}/\sqrt{\mathcal{N}_{S}+\mathcal{N}_{B}}$\\ 
\hline 
1 & 2269 & 0.45   & 412 & 10748  & 3.9 \\ 
2 & 3137 & 0.42   & 596 & 14200  & 4.9 \\ 
3 & 3595 & 0.50   & 635 & 10957  & 5.9 \\ 
4 & 2726 & 0.52   & 504 & 11514  & 4.6 \\ 
5 & 2313 & 0.51   & 404 & 8880   & 4.2 \\ 
6 & 2138 & 0.58   & 385 & 9871   & 3.8 \\
7 & 1528 & 0.55   & 278 & 6823   & 3.3 \\
\hline \hline 
$\mc{N}_{\textrm{SM}}$ & 5117800 & - & - & - & - \\ 
\hline \hline
\end{tabular} 
\caption{Total number of signal events are $\mathcal{N}_S^{bc}$ (including $1 J_V$ and $2 J_V$ topologies as shown in Table~\ref{tab:event}) and
with number of background events $\mc{N}_{\textrm{SM}}$ before $\mathrm{BDT}_{opt}$ cut. The number of signal and background events after the $\mathrm{BDT}_{opt}$ cut are
denoted by $\mathcal{N}_S$ and $\mathcal{N}_{B}$ respectively.}
\label{tab:BDT}
\end{table}

\section{Results and Discussion}
\label{results}
%%%%%%%%%%%%%%%%%%%%%%%%%%%%%%%%%%%%%%%
The IDM is a simple theoretical framework with rich phenomenology providing possible DM candidates. 
We classify the model space in four categories depending on the masses of the scalars in the model as summarized
in Table~\ref{table:Param}. 
Some of them are quite interesting in view of the observed properties of the $Z$-boson, Higgs and DM, together with fulfilling all the available theoretical constraints and from the low energy experiments.
All such constraints on the IDM are critically analyzed to establish that a hierarchical BSM spectrum with a light DM 
($m_{\rm DM} \lesssim 80$ GeV) provides an appealing scenario, as it fulfills the full observed relic density. 
Furthermore, additional constraints from the Higgs invisible decay and the DM direct detection limits leave us 
with little allowed parameter space left to be explored at the LHC, albeit a rather difficult region to explore.

Exploiting the fact that after production, the heavy BSM scaler essentially decays into boosted vector bosons 
together with light DM candidates, we propose a search strategy of a scenario consisting of two boosted fatjets with 
large MET. Hadronic decay from such boosted vector bosons carries distinctive substructures characteristically different from
the single prong large radius QCD jets and can be distinguished with moderate efficiencies using jet substructure
observables. 

It turns out that our signal of boosted $2 J_V+\slashed{E}_T$ also gets significant contributions from 
single heavy scalar productions with light DM, where the other second $J_V$ is mimicked by a QCD jet, especially since the later production is roughly one order of magnitude higher than the two $J_V$ processes.
Essentially the di-fatjet signal, after our selection cuts, turns out to be a hybrid of di-$J_V$ and mono-$J_V$ signals. The corresponding background to the mono-$J_V$ channel is also very large, which contributes to the overall background. The $V+jets$ SM processes are the dominant backgrounds to the above signal, and the sheer magnitudes of these backgrounds of order $\sim 1000$ pb make it very difficult to search for the BSM scalars of the IDM  
in any channel. We use intuitive application of jet substructure variables like the fatjet mass ($M_J$) and the $N$-subjettiness ($\tau_{21}$) which encode the internal structure of the fatjets. 

Even with these variables, it is extremely difficult to overcome the huge background and therefore, the best case cut-based analysis discovery potential remains restricted to less than $3\sigma$. While cuts on these variables, 
as detailed in Tables~\ref{tab:cut_flow_sig} and \ref{tab:cut-flow-bg}, can bring down the background to less than the $1\%$ level from the generated ones simultaneously bringing down the signal numbers to $10\%-20\%$. In the end, we do not obtain any significant improvement in the discovery potential to make it cross the desired $5\sg$ barrier for discovery.
The best LHC sensitivity is obtained for the BP3 with $m_{H^{\pm}} \approx m_A \sim 350$ GeV and significance decreases both sides of the spectrum. With the increase of $m_{H^{\pm}},m_A$, we get a higher boost for the decaying vector bosons, resulting in better discrimination power of the jet substructure variables. On the other hand, the presence of heavier particles leads to the suppressed signal cross section. Therefore, the best signal to background sensitivity is obtained only in an intermediate mass range. 

To improve the LHC discovery potential, an MVA is undertaken where we employ a total eight kinematic variables which try to devise a boosted decision tree and provide the optimum separation between 
signal and background. Instead of the rectangular cuts used in CBA, MVA can use the full potential of jet substructure variables to study the full hierarchical parameter space of the IDM which is allowed after imposing all the theoretical and experimental constraints.
The LHC sensitivity is improved to $5.6\sigma$ for BP3 using MVA. 
Hence, much of the parameter space in a well motivated scenario within the IDM framework which provides a hierarchical BSM spectrum with light DM ($m_{\rm DM} \lesssim 80$ GeV), along with an almost degenerate heavy charged Higgs and a pseudoscalar $A$ within the mass range between 250 - 550 GeV, can be excluded with $1100$ fb$^{-1}$ integrated luminosity at the 14 TeV LHC.

\acknowledgments
 The work of AB and PK is supported by Physical Research Laboratory (PRL), Department of Space, Government of India and the computations were performed using the HPC resources (Vikram-100 HPC) at PRL. TM is financially supported by the Royal Society of Arts and Sciences of Uppsala as a researcher at Uppsala University. Part of the work done by SS is supported by the DS Kothari postdoctoral fellowship granted by the UGC. 

%%%%%%%%%%%%%%%%%%%%%%%%%%%%%%%%%%%%%%%%%%%%%%%%%%%%%%
%%%%%%%%%%%%%%%%%   References   %%%%%%%%%%%%%%%%%%%%%
%%%%%%%%%%%%%%%%%%%%%%%%%%%%%%%%%%%%%%%%%%%%%%%%%%%%%%

\bibliographystyle{apsrev4-1}
\bibliography{ref}

%%%%%%%%%%%%%%%%%%%%%%%%%%%%%%%%%%%%%%%%%%%%%%%%%%%%%%

\end{document}